\documentclass[12pt,preprintnumbers,nofootinbib,amsmath,amssymb]{revtex4}
\usepackage{amsmath,amssymb,graphics,epsfig,subfig}
\usepackage{color}
\usepackage{multirow}
\usepackage{booktabs}
\usepackage{changes}
\usepackage{footnote}
\usepackage{mathrsfs}

\usepackage{hyperref}
\hypersetup{colorlinks=true,linkcolor=blue,citecolor=magenta}

\begin{document}
\renewcommand{\baselinestretch}{1.15}
\captionsetup[figure]{name=
{Fig.},labelsep=period,singlelinecheck=on,justification=raggedright}

\title{Frozen Bardeen-Dirac stars and  light ball}

\preprint{}

\author{Long-Xing Huang, Shi-Xian Sun, and Yong-Qiang Wang\footnote{E-mail: yqwang@lzu.edu.cn, corresponding author}}

\affiliation{$^{1}$ Lanzhou Center for Theoretical Physics, Key Laboratory of Theoretical Physics of Gansu Province, School of Physical Science and Technology, Lanzhou University, Lanzhou 730000, China\\
 $^{2}$ Institute of Theoretical Physics $\&$ Research Center of Gravitation, Lanzhou University, Lanzhou 730000, China}


\date{\today}

\begin{abstract}
    In this paper, we study solutions of a static spherically symmetric system, which is composed of the coupling with the Bardeen action and two Dirac fields. For the case where only the Bardeen action is present, the magnetic charge $q$ can be infinite, then when the magnetic charge is greater than a certain value $q_b$, there exists a black hole solution, which is called the Bardeen black hole (BBH). However, if the Dirac field is introduced, we find that the magnetic charge can only be smaller than the critical value $q_b$, in which there is no black hole solution. Moreover, in the region $q<q_b$, we find that if the magnetic charge exceeds another critical value $q_c$ (i.e., $q_c<q<q_b$), the frequency of the Dirac field can approach zero, and the solution where a critical horizon appears is similar to an extremal black hole outside the critical horizon but has a nonsingular interior. The Dirac fields are also almost concentrated within it. In fact, this is a frozen star solution, we call such solutions frozen Bardeen-Dirac stars (FBDSs). We analyze the light rings of FBDSs and find that there exists a ``true" light ring outside the critical horizon, but inside it, the velocity of photons is very close to zero, which leads to the formation of a ``light ball" inside the critical horizon.
\end{abstract}


\maketitle
\newpage

\section{INTRODUCTION}\label{intro}

Much of the available observational evidence suggests that there are extremely compact and massive objects in the universe~\cite{KAGRA:2021duu}. These entities surpass the compactness of neutron stars, currently recognized as the most compact directly observable objects. For instance, Sagittarius A*, the bright radio source at the center of the Milky Way, has a mass of about $4.1 \times 10^6 M_{solor}$ (where $M_{solor}$ is the mass of solar) with the radius no larger than 6.25 light-hours (6.7 billion kilometers) ~\cite{Ghez:2003qj,Ghez:2008ms}, while the mass of neutron stars below 2.2 $M_{solor}$~\cite{Rhoades:1974fn,Kalogera:1996ci,NANOGrav:2019jur,Kunz:2022wnj} and the radius above $9.27 \times 10^{-9}$ light-hours (10 - 20 kilometers). Currently, from well-established physical models that match these observational data, black holes are the most promising theoretical candidate. Therefore, Sagittarius A*, and other similarly compact objects, are often considered to be black holes~\cite{Herdeiro:2015gia}. In recent years, the event of the binary black hole merger - GW190521 - detected by LIGO and Virgo via gravitational waves~\cite{LIGOScientific:2016aoc}, as well as the Event Horizon Telescope (EHT) study of black hole shadows~\cite{Falcke:1999pj, EventHorizonTelescope:2019dse, EventHorizonTelescope:2019uob, EventHorizonTelescope:2019jan, EventHorizonTelescope:2019ths, EventHorizonTelescope:2019pgp, EventHorizonTelescope:2019ggy} further strengthen the belief in the existence of black holes. However, in general relativity, singularities are unavoidable in most physically relevant black hole solutions. But since matter cannot be infinitely compressed, spacetime singularities cannot exist in the real physical world~\cite{Einstein:1939ms}. Furthermore, singularities violate the unitarity in the quantum realm, there is an expectation that a theory of gravity that correctly combines quantum theory will solve the singularity problem. Nevertheless, this anticipation remains unrealized.

In addition, many other attempts have been made, such as the model of regular black holes without singularities~\cite{Lan:2023cvz}. The original idea about regular black holes was proposed in 1966 by A. D. Sakharov~\cite{Sakharov:1966aja} and E. B. Gliner~\cite{Gliner:1966}, who pointed out that the essential singularities can be avoided if a vacuum-like medium with the de Sitter metric was used instead of vacuum. Later, J. M. Bardeen proposed the first regular black hole model ~\cite{Bardeen:1968}, which is also called the Bardeen black hole. Initially, it was constructed by simply replacing the mass in a Schwarzschild black hole with a function dependent on $r$. Until the 2000s, E. Ayon-Beato and A. Garcia~\cite{Ayon-Beato:1998hmi} reinterpreted the Bardeen black hole model as the gravitational field of a nonlinear magnetic monopole with a large charge. And when the magnetic charge is small, solutions of this model describe the spherical soliton.  

However, one might ask what the situation of a regular black hole would be if some other matter field existed. In recent decades, the scalar field not only works well as an extension of string theory or the standard model of particle physics but can also be considered as a proxy to realistic matter such as the ideal fluid~\cite{Herdeiro:2015gia}. Thus, in gravitational physics, one of the simplest types of ``matter" that is often considered is the scalar field. The soliton solutions formed by the complex scalar field under their own gravitational are also known as the boson stars~\cite{Wheeler:1955zz, Schunck:2003kk, Liebling:2012fv,Herdeiro:2021mol}. Their dynamics can matched to observed gravitational wave signals ~\cite{CalderonBustillo:2020fyi, CalderonBustillo:2022cja} and they have been hypothesized as potential constituents of dark matter ~\cite{Lee:1995af, Suarez:2013iw, Eby:2015hsq, Chen:2020cef}.

In a more recent study, Ref.~\cite{Wang:2023tdz} investigated a model coupled to the complex scalar field and the Bardeen action. Such solutions are called Bardeen-boson stars (BBSs). It can be found that the inclusion of the scalar field restricts the
magnetic charge $q$ to values less than a certain magnetic charge $q_b$. and among solutions where $q<q_b$, no black hole solution was found. Moreover, when $q$ exceeds a critical value $q_c$ which is smaller than $q_b$, the frequency of BBSs can approach zero. As the frequency approaches zero, the metric component $g^{rr}$ is very close to zero at a critical position $r_{cH}$. Meanwhile, $g_{tt}$ is also very close to zero in $r\leq r_{cH}$. Since at $r_{cH}$, $g^{rr}$ is only close to zero but not zero, there is no event horizon, and $r_{cH}$ is merely a critical horizon radius (therefore, we refer to the spherical surface $r=r_{cH}$ as the ``critical horizon"). The scalar field is almost concentrated in its interior ($r\leq r_{cH}$). And outside this critical horizon radius ($r<r_{cH}$), the spacetime of these BBSs with frequencies approaching zero closely resembles that of an extreme black hole. These solutions are the frozen stars which were proposed by Y. B. Zel’dovich and I. D. Novikov~\cite{zeldovichbookorpaper}. The original use of the term ``frozen" star was because, from a distant viewer, the collapse of an ultra-compact object would take place over a very long period of time as if the star were frozen in place ~\cite{Ruffini:1971bza}. In addition, various interesting studies on the frozen star have been carried out~\cite{Vachaspati:2007fc,Schmelzer:2010zq,Lemos:2010te,Zhang:2010vh,Vachaspati:2016hya,Kastor:2016cqs,Brustein:2023hic,Zeng:2023ueq}.

Apart from the scalar field in Ref.~\cite{Wang:2023tdz}, can the introduction of other material fields also obtain a frozen star solution? Inspired by this, we consider the Bardeen-Dirac stars (BDSs) constituted by the coupling of two Dirac fields with the Bardeen action. In particular, this model also represents a Dirac star when the nonlinear electromagnetic field is not present. Dirac stars are particle-like solutions satisfying the Einstein-Dirac equation~\cite{Leith:2020jqw} and are also referred to as Dirac Stars, Einstein-Dirac solitons, or fermion stars~\cite{Leith:2021urf}. Ever since the pioneering work of F. Finster, J. Smoller, and S. T. Yau in 1999, who presented the first exact numerical solution for Dirac stars in spherically symmetric configurations~\cite{Finster:1998ws}, a multitude of extended models of the Dirac star have been investigated~\cite{Dzhunushaliev:2018jhj, Dzhunushaliev:2019kiy, Dzhunushaliev:2019uft, Leith:2022icf, Dzhunushaliev:2022wnd, Liang:2023ywv, Sun:2023bdh}. All of these models exhibit numerous properties akin to those found in the boson star~\cite{Herdeiro:2017fhv, Herdeiro:2019mbz, Herdeiro:2020jzx}. In this work, we find that similar to the case of the Bardeen-boson star, the inclusion of the Dirac field renders finding a black hole solution impossible. Moreover, at frequencies close to zero, we obtain solutions for what we call frozen Bardeen-Dirac stars. By analyzing the effective potential of FBDSs, we find that there are two light rings in these FBDSs. However, considering that the velocity of photons inside the critical horizon is nearly zero, it seems to be a better term to refer to the inside ``light ring" as ``light ball".

The paper is organized as follows. In Sec.~\ref{sec: model}, we describe the model of BDSs and its two global charges. In Sec.~\ref{sec: boun}, the boundary conditions of the Bardeen-Dirac stars are obtained. numerical results of the Bardeen-Dirac stars and analysis of their physical properties are present in Sec.~\ref{sec: num}. The conclusion and discussion are given in Sec.~\ref{sec: conclusion}.

\section{MODEL}\label{sec: model}
 We consider Einstein’s gravity minimally coupled two Dirac fields and the nonlinear electromagnetic field of the Bardeen model (For a spherically symmetric configuration, one should consider (at least) two spinors, with the equal mass $\mu$). The model has the following action to describe:
    	\begin{equation}
			\mathcal{S}=\int d^4 x \sqrt{-g}\left[\frac{R}{16 \pi G}+\mathcal{L}^{B}+\mathcal{L}^{D}\right], \label{eq:action}
		\end{equation}	
  with the Lagrangian densities of electromagnetic field~\cite{Ayon-Beato:2000mjt} and the Dirac field are
		\begin{equation}
		    			\mathcal{L}^{B}=-\frac{3}{2s}(\frac{\sqrt{2q^2\mathcal{F}}}{1+\sqrt{2q^2\mathcal{F}}})^{5/2},  
         			\label{eq:lagrangianb}
		\end{equation}      
  	\begin{equation}
			\mathcal{L}^D=-i \sum_{k=1}^2\left[\frac{1}{2}\left(\hat{D}_\mu \bar{\Psi}^{(k)} \gamma^\mu \Psi^{(k)}-\bar{\Psi}^{(k)} \gamma^\mu \hat{D}_\mu \Psi^{(k)}\right)+\mu \bar{\Psi}^{(k)} \Psi^{(k)}\right].
			\label{eq:lagrangiandd}
		\end{equation}  
	%
Where $G$ represents the Newtonian gravitational constant, $R$ is the Ricci scalar, $\mathcal{F}=\frac{1}{4}F_{\mu \nu} F^{\mu \nu}$ with the electromagnetic field $F_{\mu \nu}=\partial_{\mu}A_{\nu}-\partial_{\nu}A_{\mu}$, in which $A_{\mu}$ is the electromagnetic $4$-potential. And here, $\Psi^{(k)}$ ($k = 1$, $2$) denotes the Dirac field. $q$, $s$, and $\mu$ are three independent parameters, where $q$ represents the magnetic charge and $\mu$ denotes the Dirac field mass. $\gamma^{\mu}$ are gamma matrices of curved spacetime, which satisfy the anticommutation relations $\{\gamma^{\mu},\gamma^{\nu}\}=2g^{\mu \nu}$. Moreover, $\Bar{\Psi}^{(k)}\equiv\Psi^{(k)\dagger}\xi$ denote sthe Dirac conjugate, here $\Psi^{(k)\dagger}$ is the Hermitian conjugate of the Dirac field. For representations of the Hermitizing matrix $\xi$, we choose $\xi=-i\gamma^0$~\cite{Dolan:2015eua}. $\hat{D}_\mu=\partial_\mu+\Gamma_\mu$ is the spinor covariant derivative, where $\Gamma_\mu$ denotes the spin connection matrices. 

It is worth noting that there are two special cases of action (\ref{eq:action}). Firstly, when $q = 0$ and the Dirac field does not vanish, action (\ref{eq:action}) describes the Einstein-Dirac theory. Second, when the Dirac field vanishes and $q \neq 0$, the model degenerates into the Bardeen theory which describes asymptotically spherical solitons and black holes, and the metric of the Bardeen theory is expressed analytically in standard spherical coordinates $(t,r,\theta,\varphi)$  as
\begin{equation}
    ds^2=-f(r)dt^2+f(r)^{-1}dr^2+r^2\left(d\theta^2+\sin^2\theta d\varphi^2\right),
\end{equation}
where
\begin{equation}
    f(r)=1-\frac{q^3r^2}{s(r^2+q^2)^{3/2}}.
\end{equation}
The asymptotic behavior for this metric function $f(r)$ at infinity is given by
\begin{equation}
    f(r)=1-2\frac{q^3}{2s}/ r+O(1/r^3).
\end{equation}
From the term $1/r$, one can deduce the ADM mass $M$ of this configuration as $M = q^3/2s$. Furthermore, the equation $f(r) = 0$ has real roots only when $q \geq q_b = 3^{3/4}\sqrt{s/2}$, therefore, the solution for Bardeen spacetime with $q \geq q_b$ corresponds to a black hole with the event horizon, and in which the solution with $q=0$ is an extremal Bardeen black hole solution. Conversely, when $q < q_b$, the equation has no event horizon, therefore, there is no black hole solution.

For the general case, varying the action (\ref{eq:action}) with respect to the metric, electromagnetic field, and Dirac fields, we derive the following equations:
\begin{equation}
    R_{\mu \nu} - \frac{1}{2}g_{\mu \nu}R - 8\pi G (T^{B}_{\mu \nu}+T^{D}_{\mu \nu})=0,
    			\label{eq:einstein}
\end{equation}
\begin{equation}
    \nabla_a\left(\frac{\partial \mathcal{L}^{B}}{\partial \mathcal{F}} F^{a b}\right)=0,
    			\label{eq:equationb}
\end{equation}
\begin{equation}
    \gamma^\mu \hat{D}_\mu \Psi^{(k)}-\mu \Psi^{(k)}=0,
    			\label{eq:equationd}
\end{equation}
here, $R_{\mu\nu}$ is the Ricci curvature tensor, $T_{\mu \nu}^{B}$ and $T_{\mu \nu}^D$ represent the energy-momentum tensors for the electromagnetic field and the Dirac fields, respectively:
\begin{equation}
    T_{\mu \nu}^{B}=-\frac{\partial \mathcal{L}^{B}}{\partial \mathcal{F}} F_{\mu \sigma} F_\nu{ }^\sigma+g_{\mu \nu} \mathcal{L}^{B},
    			\label{eq:energymomb}
\end{equation}
\begin{equation}
T_{\mu \nu}^{D}=\sum_{k=1}^2-\frac{i}{4}\left(\bar{\Psi}^{(k)} \gamma_\mu \hat{D}_\beta \Psi^{(k)}+\bar{\Psi}^{(k)} \gamma_\nu \hat{D}_\mu \Psi^{(k)}-\hat{D}_\mu \bar{\Psi}^{(k)} \gamma_\nu \Psi^{(k)}-\hat{D}_\nu \bar{\Psi}^{(k)} \gamma_\mu \Psi^{(k)}\right).
    			\label{eq:energymomd}
\end{equation}
In this paper, we consider only spherically symmetric configurations. The corresponding spacetime metric is most conveniently chosen in the form
\begin{equation}
    d s^2=-n(r)o^2(r) d t^2+\frac{1}{n(r)} d r^2+r^2\left(d \theta^2+\sin ^2 \theta d \varphi^2\right).
        \label{eq:metric}
\end{equation}
The two metric functions $n(r)$ and $o(r)$ are functions of the radial variable $r$. In addition, one could introduce the following ansatz of the electromagnetic field and the Dirac field: 
\begin{equation}
    A=q \cos (\theta) d \varphi,
    \label{eq:anastzb}
\end{equation}
\begin{align}
  \Psi^{(1)}=e^{-i\left(\frac{1}{2} \varphi+\omega t\right)}\left(\begin{array}{c}
(-a(r)+i b(r)) \sin \left(\frac{\theta}{2}\right) \\
(i a(r)+b(r)) \cos \left(\frac{\theta}{2}\right) \\
(a(r)+i b(r)) \sin \left(\frac{\theta}{2}\right) \\
(-i a(r)+b(r)) \cos \left(\frac{\theta}{2}\right)
\end{array}\right), 
\label{eq:anastzd1}
\end{align}
\begin{align}
\Psi^{(2)}=e^{i\left(\frac{1}{2} \varphi-\omega t\right)}\left(\begin{array}{c}
(a(r)-i b(r)) \cos \left(\frac{\theta}{2}\right) \\
(i a(r)+b(r)) \sin \left(\frac{\theta}{2}\right) \\
(-a(r)-i b(r)) \cos \left(\frac{\theta}{2}\right) \\
(-i a(r)+b(r)) \sin \left(\frac{\theta}{2}\right)
\end{array}\right),
\label{eq:anastzd2}
\end{align}
here, $\omega$ is the frequency of the Dirac field, $a(r)$ and $b(r)$ are radial functions.

Using the metric (\ref{eq:metric}) and the ansatz (\ref{eq:metric} --~\ref{eq:anastzd2}) for the Dirac field and electromagnetic field, we can derive explicit expressions for the Dirac field equations and the Einstein field equations
\begin{equation}
 a'+a\left(\frac{o'}{2o}+\frac{n'}{4n}+\frac{1}{r\sqrt{n}}+\frac{1}{r}\right)+b \left(\frac{\mu}{\sqrt{n}}-\frac{\omega}{no}\right)=0,
\end{equation}
\begin{equation}
b'+b\left(\frac{o'}{2o}+\frac{n'}{4n}-\frac{1}{r\sqrt{n}}+\frac{1}{r}\right)+a \left(\frac{\mu}{\sqrt{n}}+\frac{\omega}{no}\right)=0,
\end{equation}
\begin{equation}
n'+\frac{n}{r}+\frac{32G\pi \omega r }{o\sqrt{n}}(a^2+b^2)+\frac{12\pi G q^5 r}{s \left(q^2+r^2\right)^{5/2}}-\frac{1}{r}=0,
\end{equation}
%
%
\begin{equation}
o'+\frac{16\pi G ro  }{\sqrt{n}}(ab'-a'b)-\frac{16\pi G r \omega}{n^{3/2}}(a^2+b^2)=0,
\end{equation}
%
where the prime denotes differentiation with respect to the radial coordinate. These four equations, being first-order ordinary differential equations, can be solved as a boundary value problem by applying suitable boundary conditions.

Moreover, the action of the Dirac field possesses invariant under the global $U(1)$ transformation $\Psi \rightarrow e^{i\alpha}\Psi$ with a constant $\alpha$. This implies the existence of a global conserved charge. According to Noether's theory
\begin{equation}
    Q=\int_\mathscr{S} j_{\mu}n^{\mu}dV,
\end{equation}
where $\mathscr{S}$ denotes a spacelike hypersurface, $n^{\mu}$ is the unit normal vector of $\mathscr{S}$, and $j_{\mu}$ is a conserved current associated with the $U(1)$ symmetry of the Dirac field
\begin{equation}
    j_{\mu}=\sum_{k=1}^2 \bar{\Psi}^{(k)}\gamma_\mu\Psi^{(k)}.
\end{equation}

 Besides the Noether charge $Q$, the ADM mass $M$ is another global ``charge" of BDSs. In asymptotically flat spacetime, with the killing vector $\zeta^{\nu}$ being timelike everywhere, the ADM mass can be determined from the Komar expressions
\begin{equation}
    M=\frac{1}{4\pi G}\int_\mathscr{S} R_{\mu \nu} n^{\mu} \zeta^{\nu} dV.
\end{equation}
 Furthermore, the ADM mass of BDSs can be determined from the asymptotic behaviors of solutions at infinity
\begin{equation}
    g_{tt}=-no^{2}=-1+\frac{2GM}{r}+\cdots.
\end{equation}
\section{BOUNDARY CONDITIONS}\label{sec: boun}
As described in the Sec. \ref{sec: model}, before numerically solving the coupled ordinary differential equations, we need to obtain the regular behavior of four functions $n(r)$, $o(r)$, $a(r)$, $b(r)$ at the origin and their asymptotic behavior at infinity. Analyzing the form of the field equations at the origin, we can derive the following boundary conditions
\begin{equation}
    n(0)=1,\quad o(0)=o_0,\quad a(0)=\partial_rb(0)=0.
\end{equation}
where $o_0$ is a finite value. Moreover, considering the assumption of asymptotic flatness of the solution, at $r\rightarrow \infty$ one imposes 
\begin{equation}
    n(\infty)=1,\quad o(\infty)=1,\quad a(\infty)=b(\infty)=0.
\end{equation}

To facilitate numerical computations, we employ the following scaling transformations to get the following dimensionless variables
\begin{equation}
    r= \mu_{no} r_{no},\quad \omega= \frac{\omega_{no}}{\mu_{no}},\quad \mu = \frac{\mu_{no}}{\mu_{no}}=1,\quad a = \sqrt{\frac{4\pi}{\mu_{no} M_{Pl}^2}}a_{no},\quad b =\sqrt{\frac{4\pi}{\mu_{no} M_{Pl}^2}}b_{no}.
\end{equation}
The label ``no" indicates physical quantities that are not dimensionless, where $M_{Pl}=1/\sqrt{G}$ denotes the Planck mass. Moreover, without loss of generality, unless specified otherwise, we set $s=0.3$, and $G=\frac{1}{4\pi}$. For other values, there are consistent conclusions. The only remaining input parameters are the frequency $\omega$ and the magnetic charge $q$ of the Dirac field. Moreover, we introduce a conformal transformation $x=r/(1+r)$ to map the semi-infinite region $\left[0,\infty\right)$ to the finite unit interval $\left[0,1\right]$.

We use the finite element method to solve a system of four first-order ordinary differential equations, the number of grid points in the integration region $\left[0,1\right]$ is 10000. We utilize the Newton-Raphson method as the iterative scheme, and the relative error for the numerical solutions in this paper is estimated to be below $10^{-5}$.
\section{NUMERICAL RESULTS}\label{sec: num}
\begin{figure}[!htbp]
		\centering		
		\subfloat[]{ 
			\includegraphics[height=.30\textheight,width=.34\textheight, angle =0]{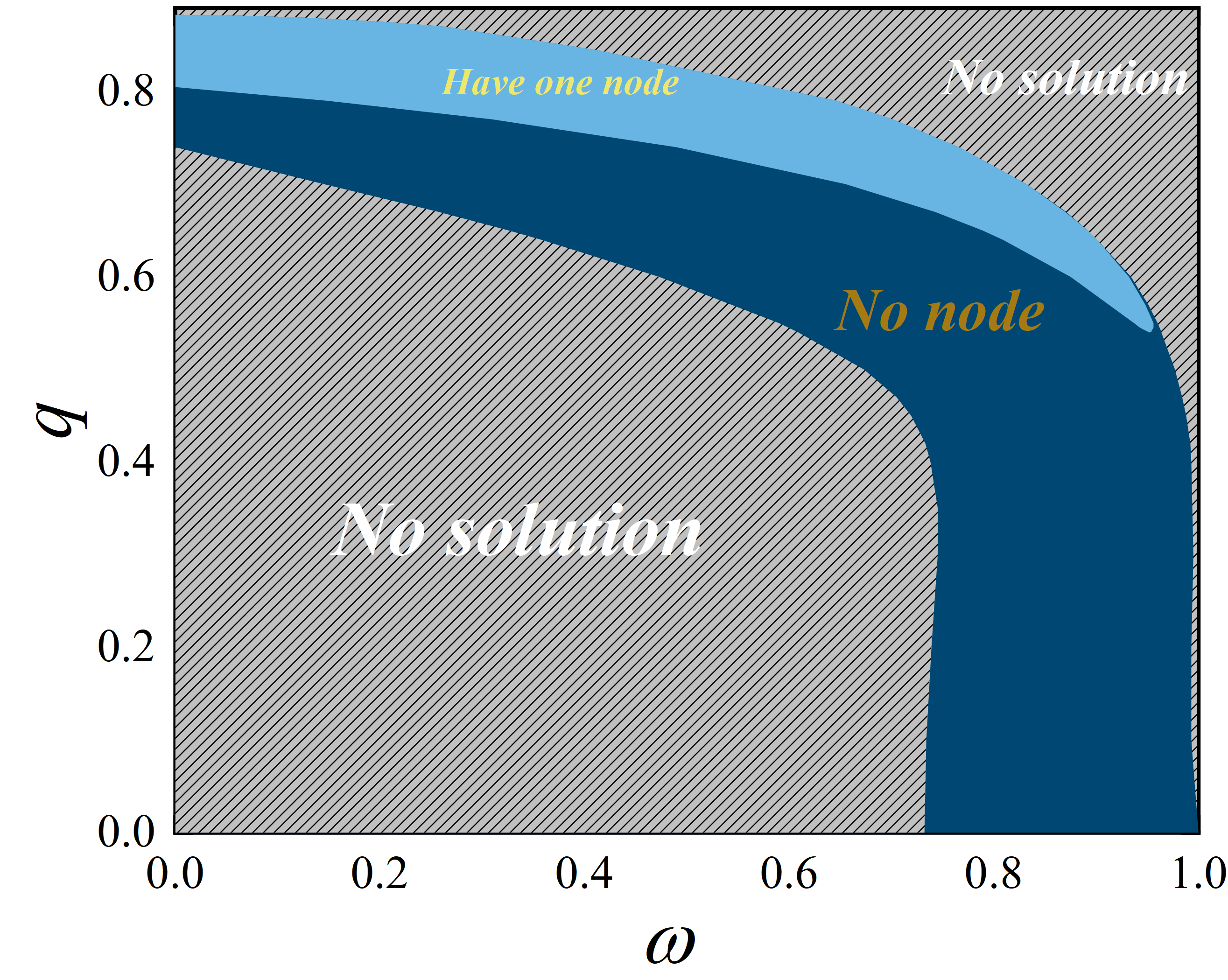}
			\label{fig:domainofexistencewq}
		}	 
  		\subfloat[]{  
			\includegraphics[height=.30\textheight,width=.34\textheight, angle =0]{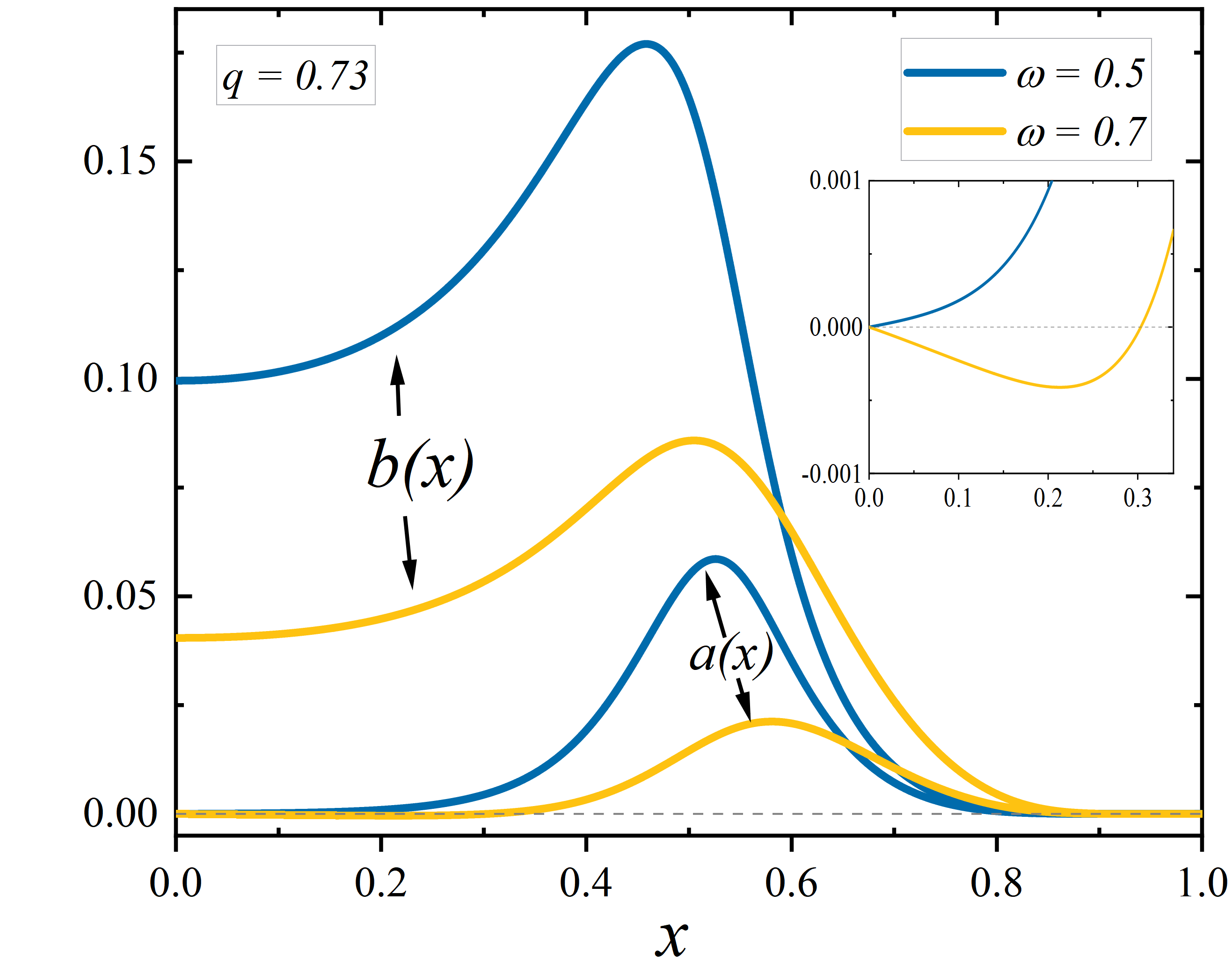}
			\label{fig:fieldfuna073}
		}	
		\caption{Left panel: The domain of existence for solutions and nodes of BDSs in a magnetic charge $q$ vs. the Dirac field frequency $\omega$. Right panel: the Dirac field functions $a$ and $b$ for BDSs with $q=0.73$.}
		\label{fig:domainofexistencewqand073}		
		\end{figure}

   In this section, we present our numerical solutions. Fig. \ref{fig:domainofexistencewq} shows the domain of existence of solutions we obtained (deep blue and light blue regions). Firstly, when the magnetic charge is greater than $q\geq 3^{3/4}\sqrt{s/2}= 3^{3/4}\sqrt{0.15}$, we cannot find solutions for BDSs. Secondly, we observe that when the magnetic charge $q$ is greater than $0.54$, and if the frequency $\omega$ of the Dirac field takes an appropriate value, the BDSs' Dirac field function $a$ will have an intersection with the $x$-axis (refer to Fig. \ref{fig:fieldfuna073} for an illustrative example), i.e., the function $a$ has a radial node. In contrast, the field function $b$ has no radial node for all solutions\footnote{It should be noted that, due to properties of numerical solutions, we consider nodes only when the absolute value of the smaller peak (refer to the inset of \ref{fig:fieldfuna073} ) of the field function is greater than $10^{-5}$.}. 
	\begin{figure}[!htbp]
		\centering	
  		\subfloat{  
			\includegraphics[height=.30\textheight,width=.34\textheight, angle =0]{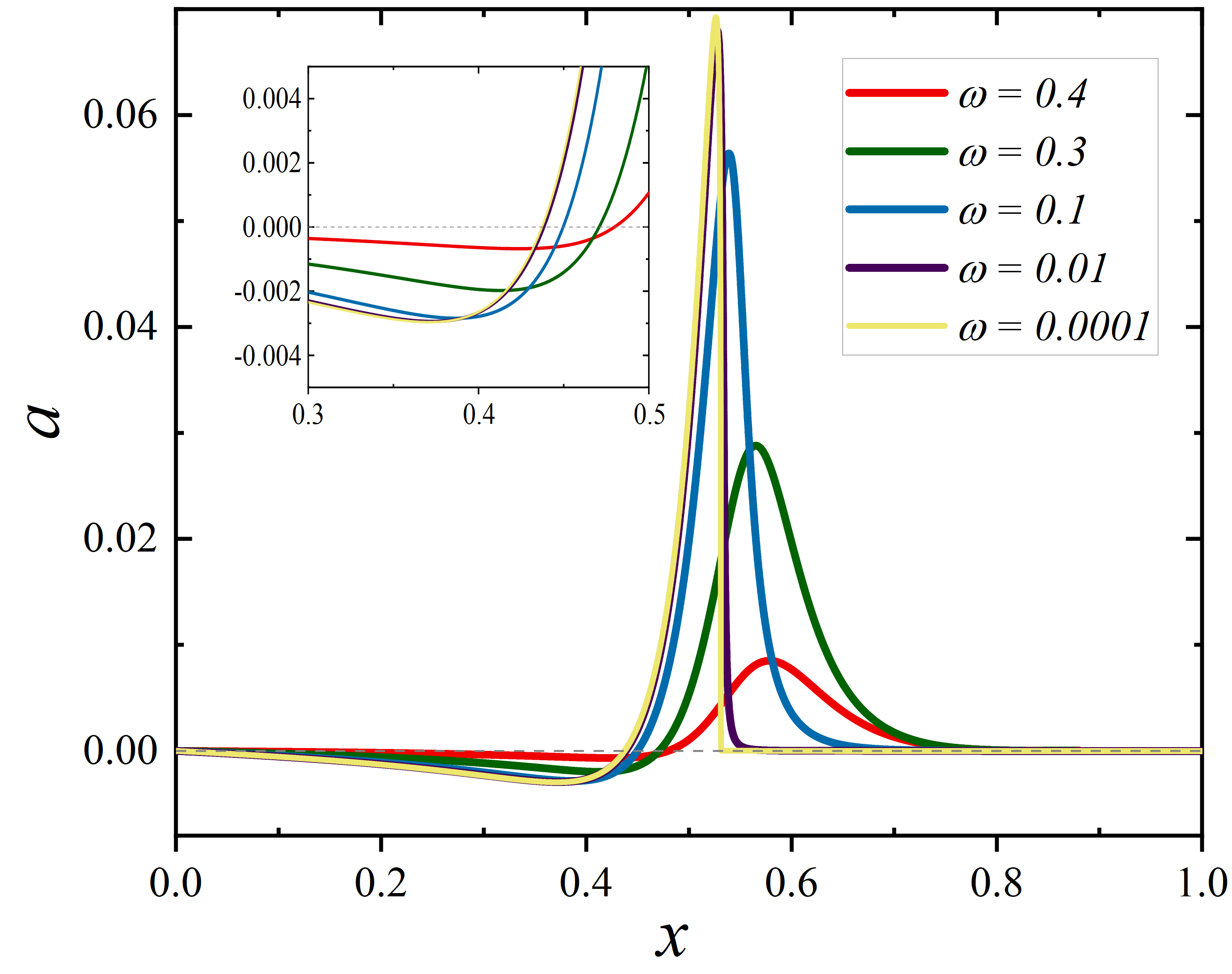}
			\label{fieldfuna}
		}
  		\subfloat{  
			\includegraphics[height=.30\textheight,width=.34\textheight, angle =0]{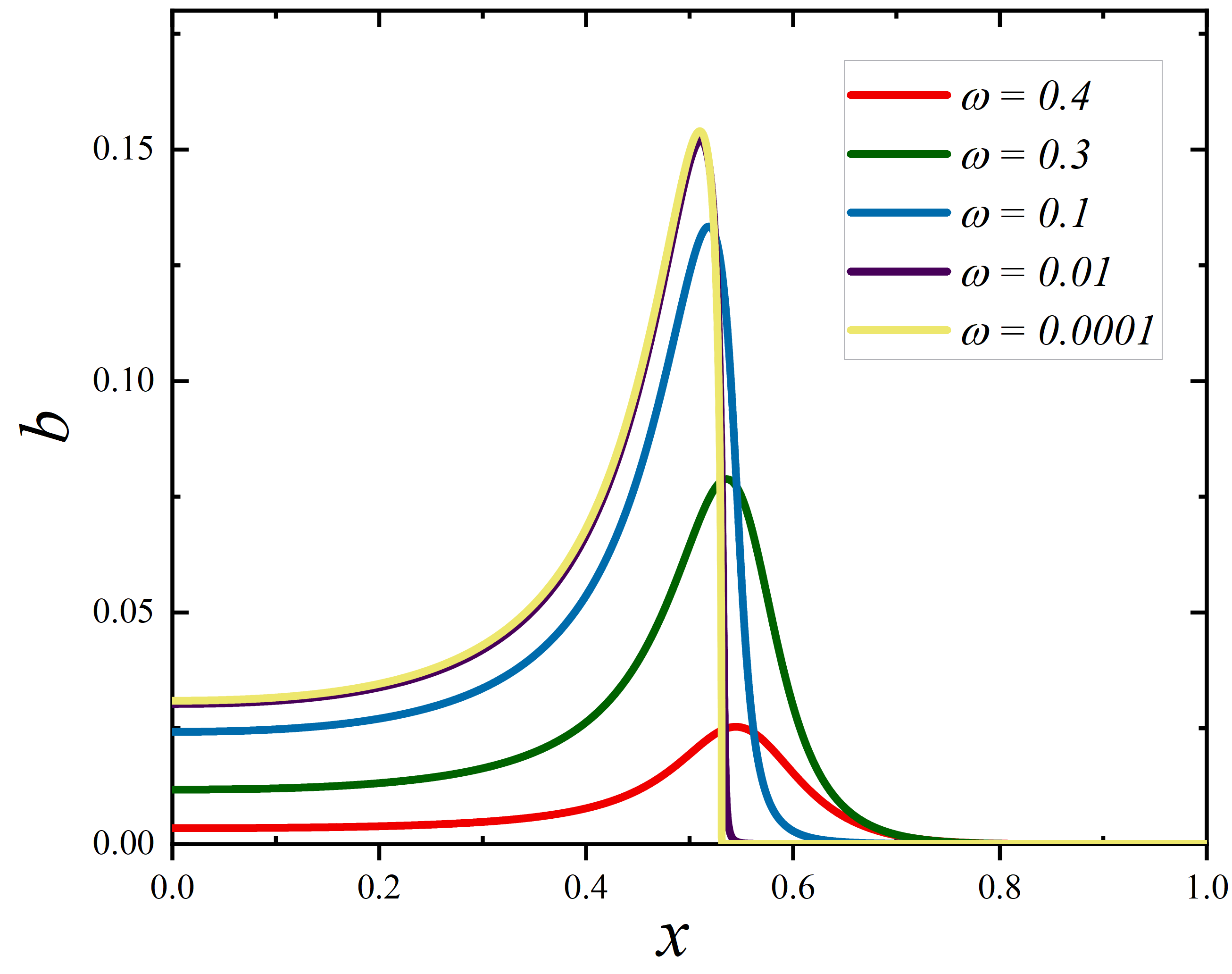}
			\label{fieldfunb}
		}
  \quad
		\subfloat{  
			\includegraphics[height=.30\textheight,width=.34\textheight, angle =0]{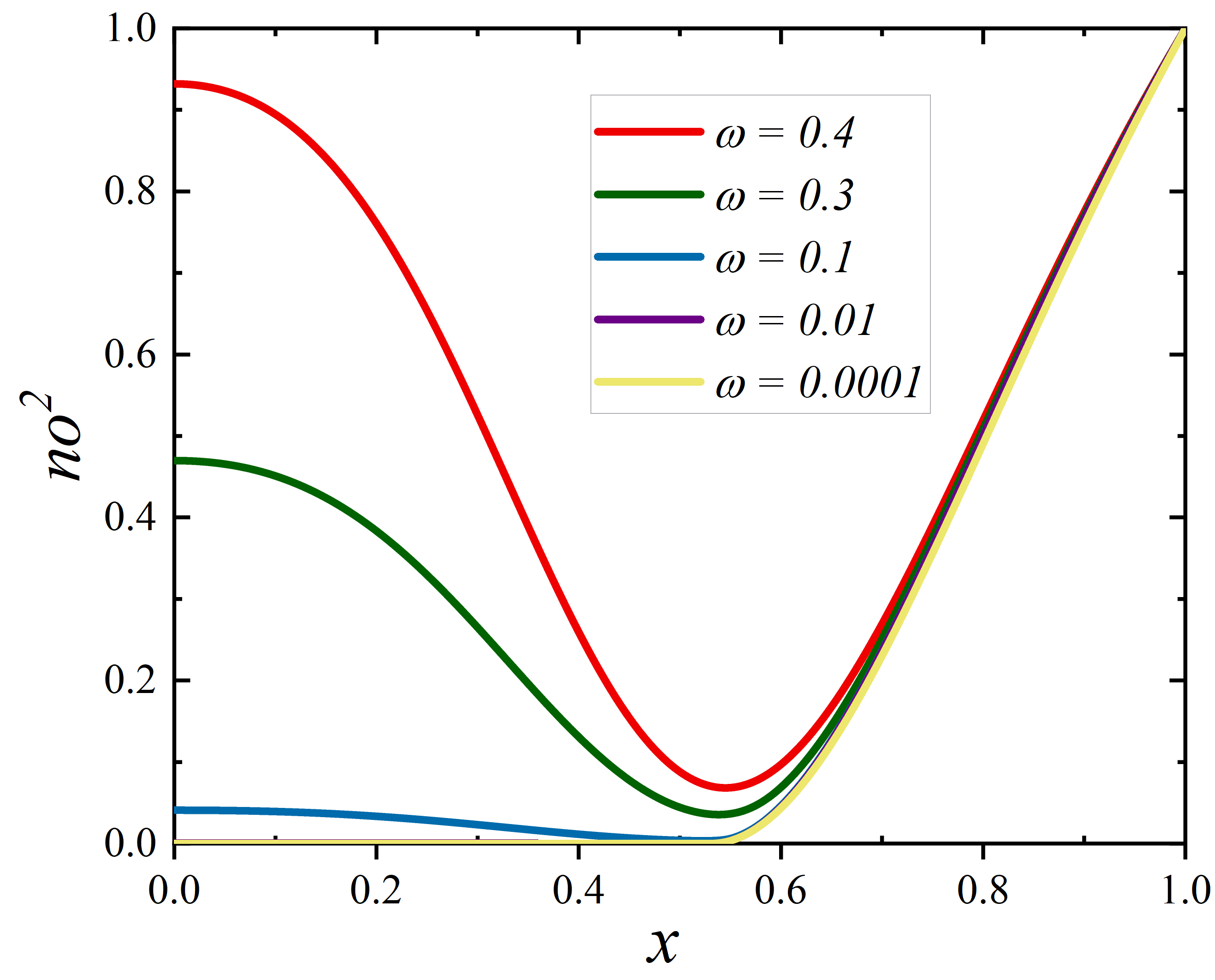}
		}
		\subfloat{
			\includegraphics[height=.30\textheight,width=.34\textheight, angle =0]{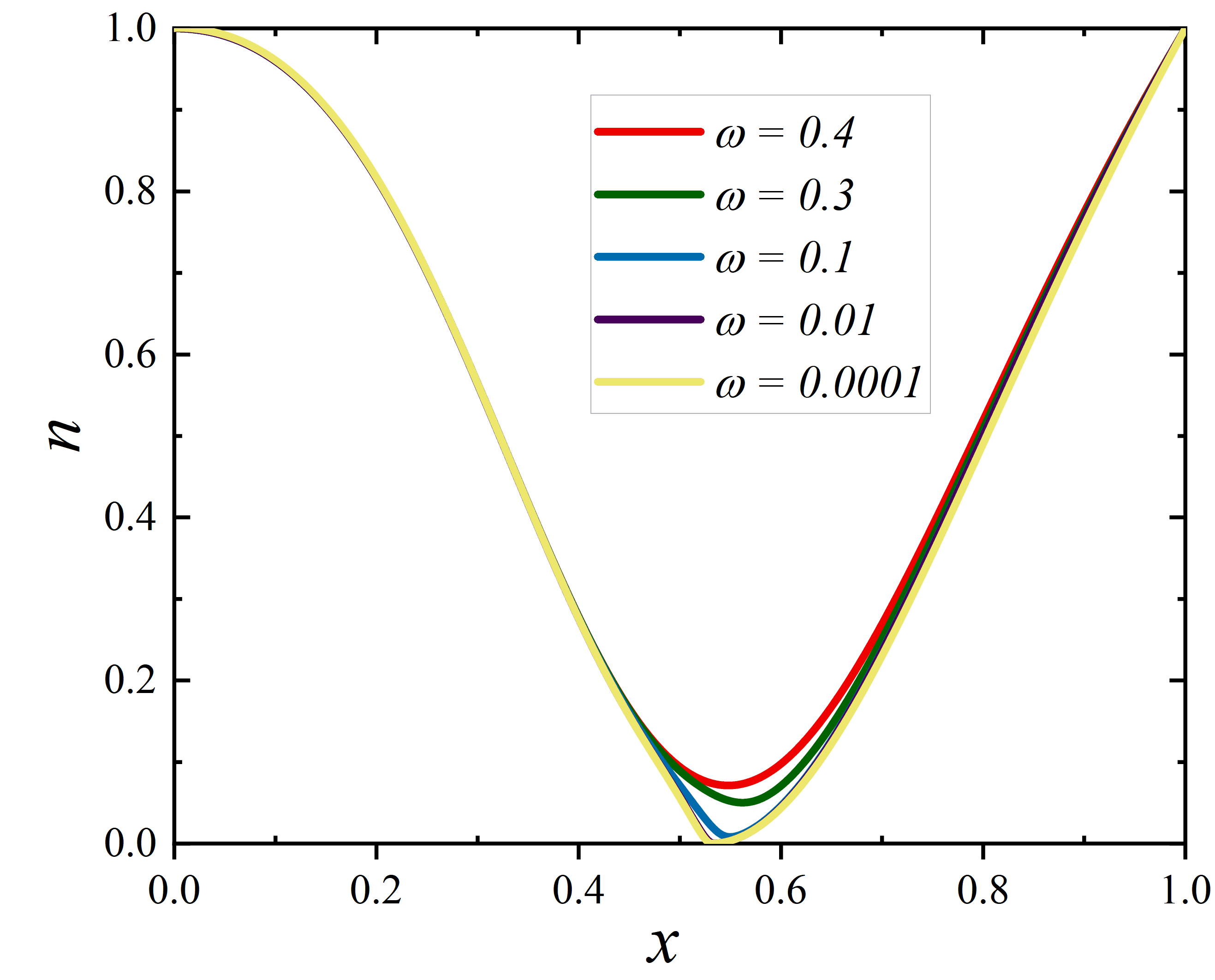}
   		} 	
       \quad
       		\subfloat{  
			\includegraphics[height=.30\textheight,width=.34\textheight, angle =0]{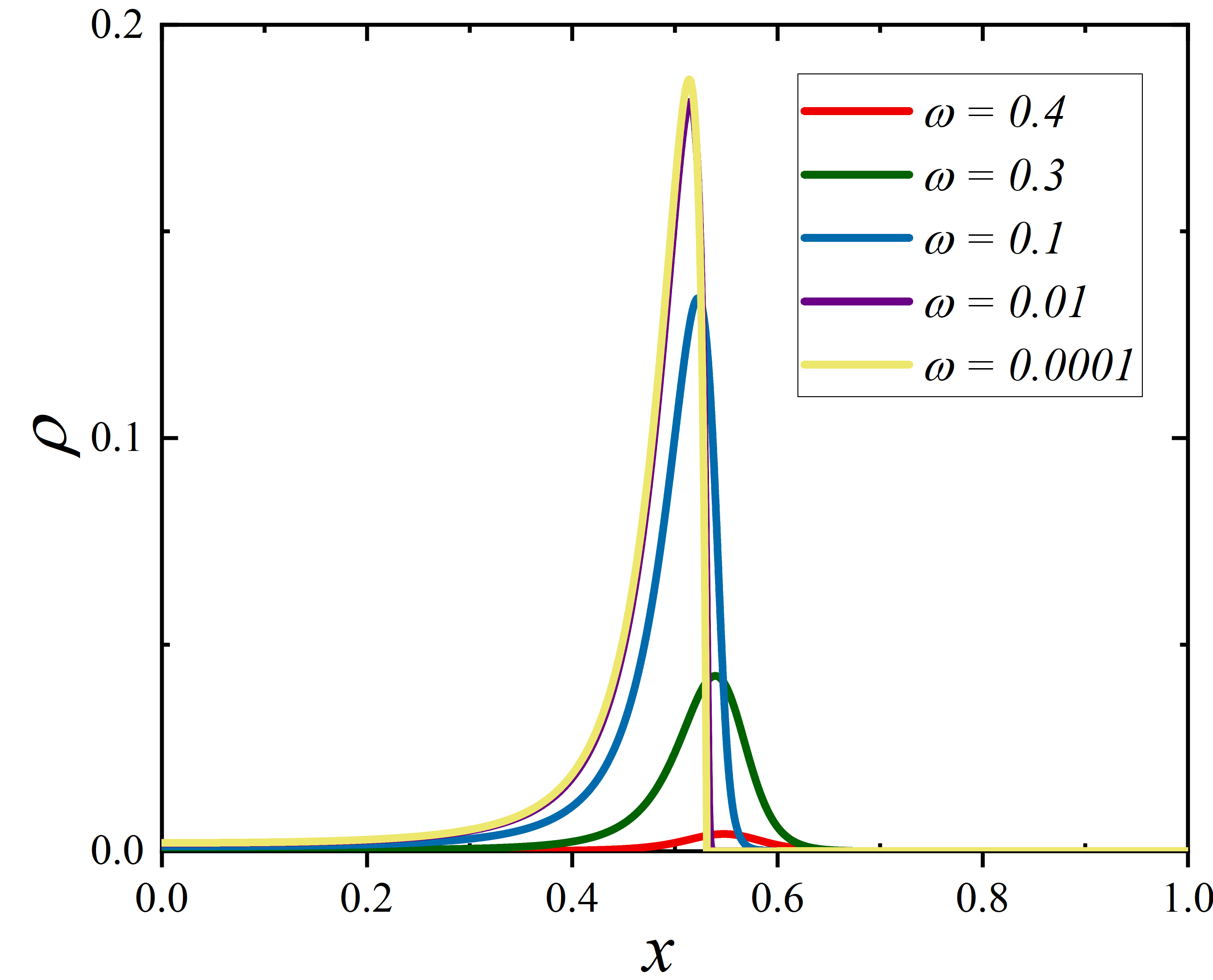}
		}
		\subfloat{
			\includegraphics[height=.30\textheight,width=.34\textheight, angle =0]{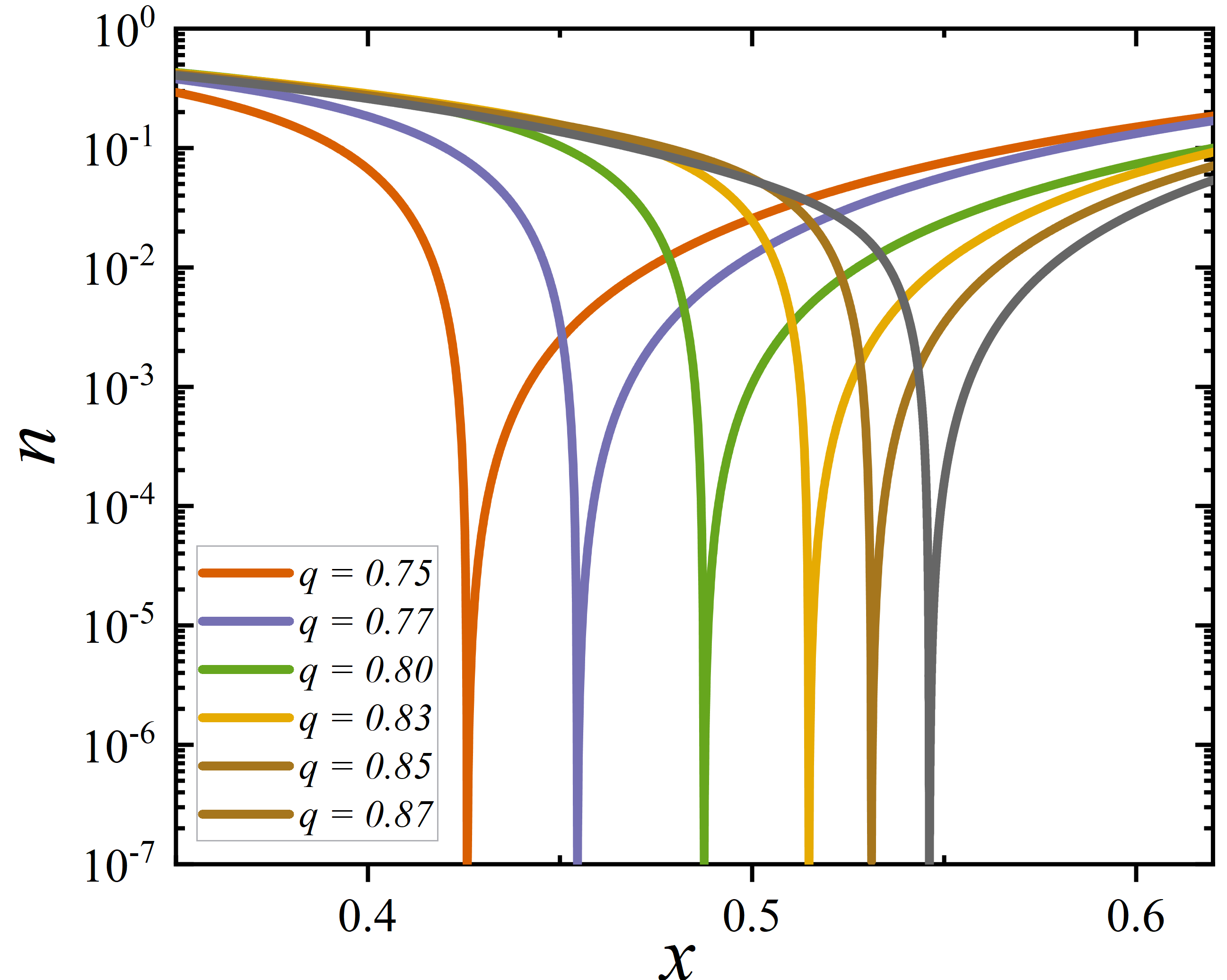}
   		} 
		\caption{The Dirac field functions $a$, $b$, the metric functions $no^2$, $n$ and the energy density $\rho$ profiles for BDSs of several frequency $\omega$, all with the same magnetic charge $q=0.85$ and $s=0.3$. The vertical axis of the bottom right panel is represented in logarithmic coordinates. }
		\label{fig:metricfunctionandrho}	
		\end{figure}
    
    Furthermore, from the top panel of Fig. \ref{fig:metricfunctionandrho}, it can be observed that when $\omega$ approaches zero\footnote{From the top panel of Fig. \ref{fig:metricfunctionandrho}, it can be observed that the solutions with a frequency of $\omega=0.1$ (indicated by the yellow line) and $\omega=0.0001$ (indicated by the purple line) almost overlap. If we were to further increase the grid resolution, the Dirac field frequency could be taken even smaller. However, this would demand a substantial amount of computational resources, and the shape and properties of the solutions would not undergo particularly noticeable changes. Therefore, in our computations, the minimum frequency of the Dirac field is $\omega=0.0001$.}, and both field functions almost converge to the same radial coordinate $x_{cH}$. Beyond this coordinate, the field functions $a(r)$ and $b(r)$ sharply decay (top panels), forming a very high wall. Moreover, as shown in the middle panel of Fig. \ref{fig:metricfunctionandrho}, inside $x_{cH}$ (where $x_{cH}=r_{cH}/(1+r_{cH})$), the metric component $-g_{tt}=no^2$   is less than $10^{-7}$, while $g^{rr}=n$ at $x_{cH}$ is also less than $10^{-7}$ - both very close to zero but have not vanished. Considering these properties are similar to the event horizon of a black hole, we refer to the sphere represented by $x=x_{cH}$ as the ``critical horizon". And the solutions under these parameters are termed ``frozen Bardeen-Dirac stars".

Moreover, in the bottom left panel of Fig. \ref{fig:metricfunctionandrho}, it can be seen that when the frequency is very small, as the radial coordinate decreases, the energy density of the Dirac field $\rho=-g^{0 \mu}T^{D}_{\mu 0 }$ instantaneously increases to a large value at the critical horizon and then rapidly decays not far away. This resembles the formation of a spherical shell. To better visualize the variation of the critical horizon, logarithmic plots (bottom right panel) of the metric component $g^{rr}=n$ for FBDSs with various magnetic charges are provided. On the one hand, it can be observed that $g^{rr}$ is below $10^{-7}$ for all solutions. On the other hand, with increasing magnetic charge, the critical horizon will move outward.

\begin{figure}[!htbp]
		\centering		
		\subfloat{  
			\includegraphics[height=.30\textheight,width=.34\textheight, angle =0]{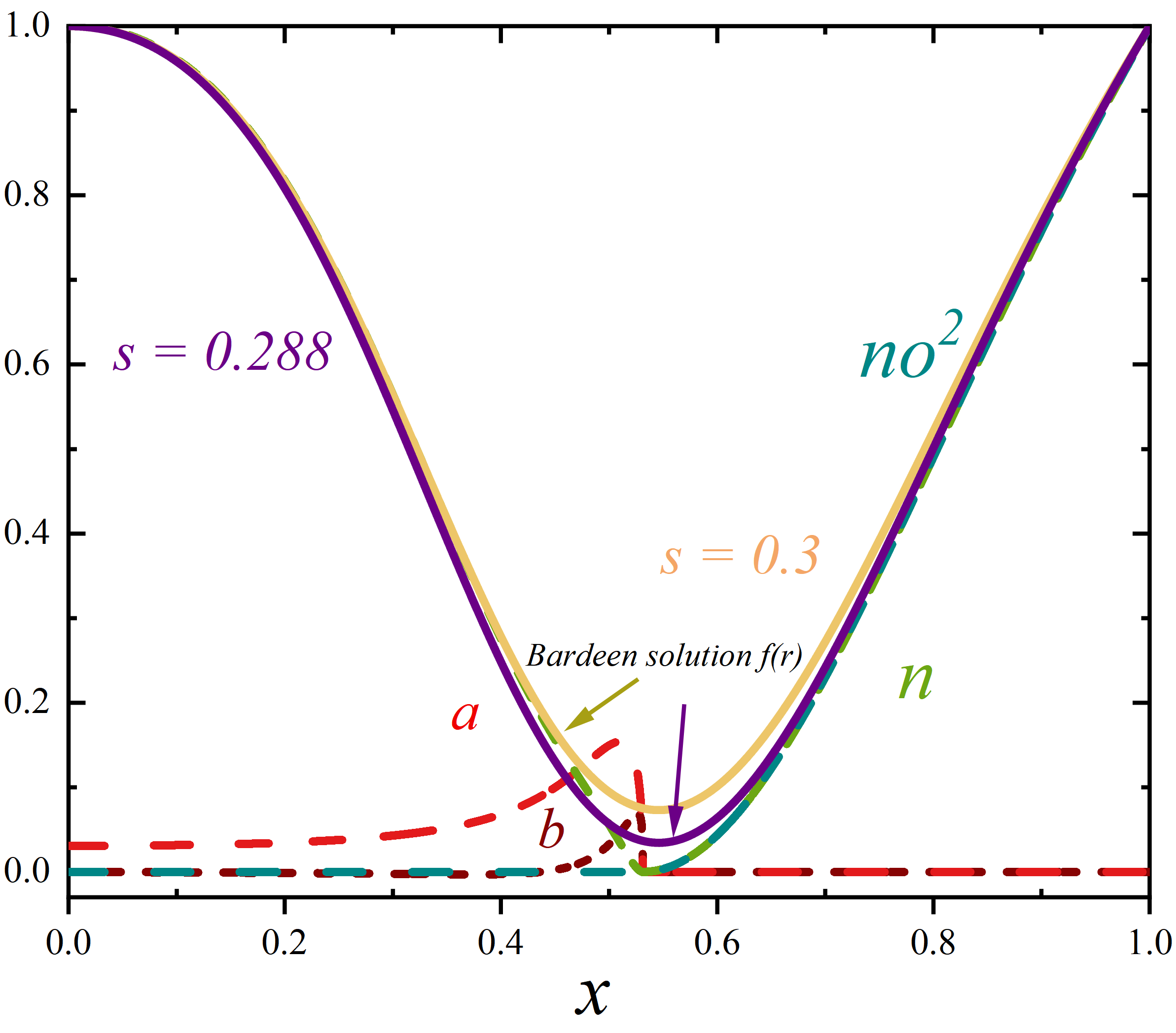}
			\label{comparebardeen}
		}	 
		\caption{The field distribution of the FBDSs with $q=0.85$, $\omega=0.85$ and $s=0.3$. The purple and orange curves denote the Bardeen solutions with $s=0.3$ and $s=0.288$ for the magnetic charges $q=0.85$.}
		\label{fig:comparebardeen}		
		\end{figure}

It is noteworthy that, similar to the extreme Bardeen-boson stars in Ref.~\cite{Wang:2023tdz}, this critical horizon arises due to the introduction of a matter field. Fig. \ref{fig:comparebardeen}, using $q=0.85$ as an example, compares the metric function $f(r)$ of the Bardeen solution (solid orange line) with $-g_{tt}=no^2$ of the FBDSs solution (dashed line). Near the critical horizon, the presence of the Dirac field causes $f(r)$ to decrease to almost zero but not exactly zero. We compared the Bardeen solutions with the same mass as this FBDS (represented by the light green dashed line). We found that the parameter $s$ needs to decrease to $0.288$ (solid purple line).

\begin{figure}[!htbp]
		\centering		
		\subfloat{  
			\includegraphics[height=.30\textheight,width=.34\textheight, angle =0]{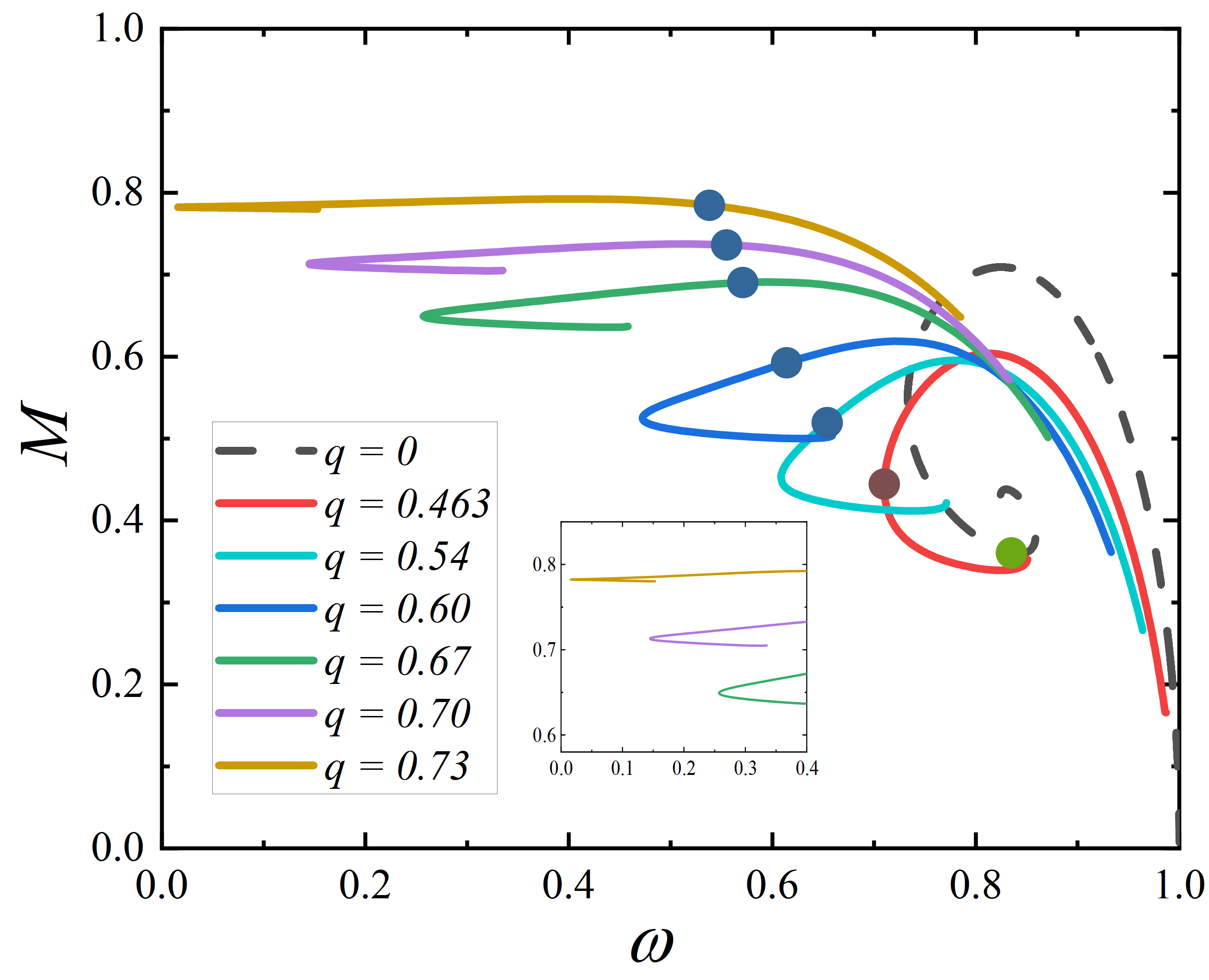}
			\label{fig:littleadm}
		}
		\subfloat{
			\includegraphics[height=.30\textheight,width=.34\textheight, angle =0]{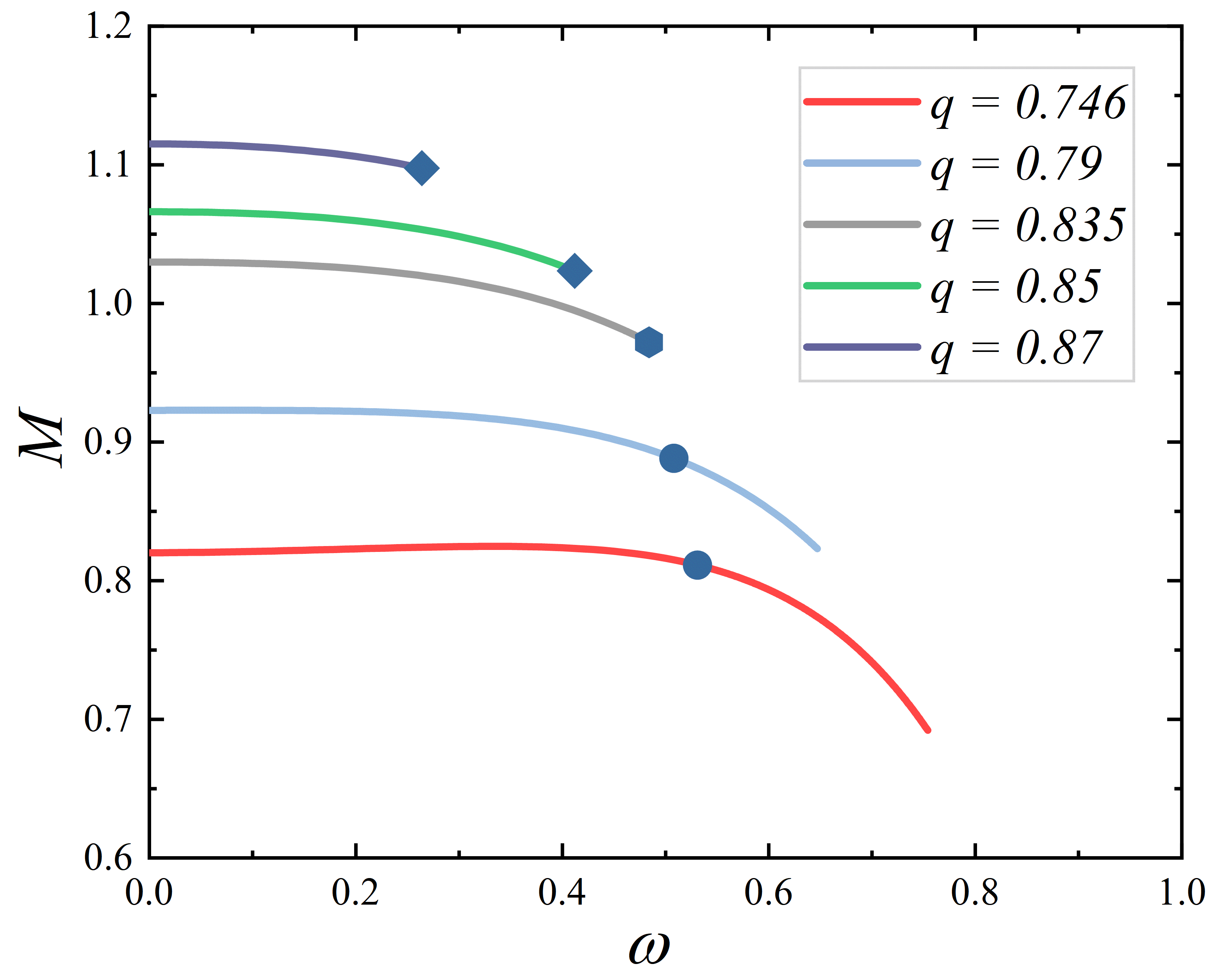}
			\label{fig:bigadm}
   		} 
   \quad
		\subfloat{  
			\includegraphics[height=.30\textheight,width=.34\textheight, angle =0]{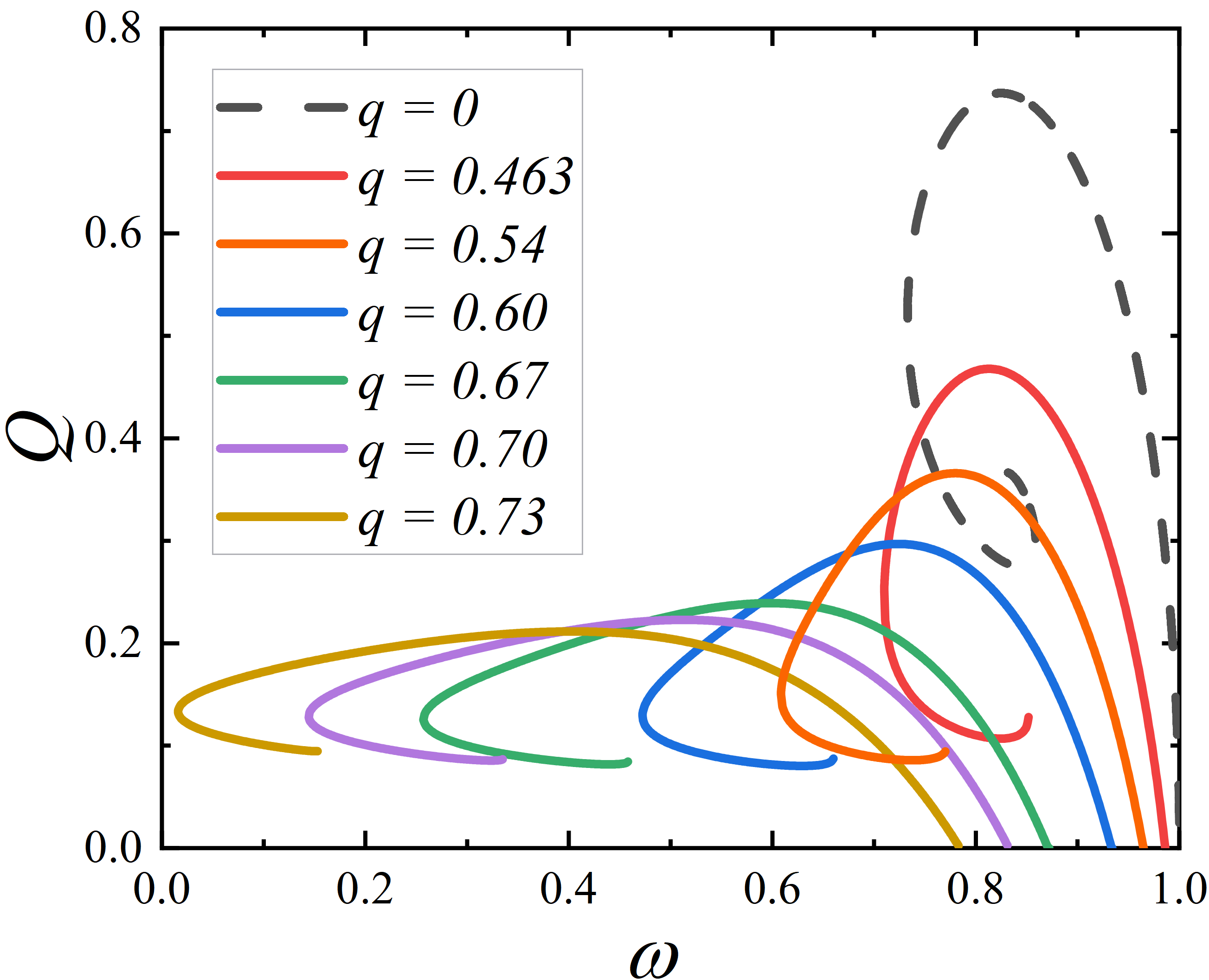}
			\label{fig:bigcharge}
		}
		\subfloat{
			\includegraphics[height=.30\textheight,width=.34\textheight, angle =0]{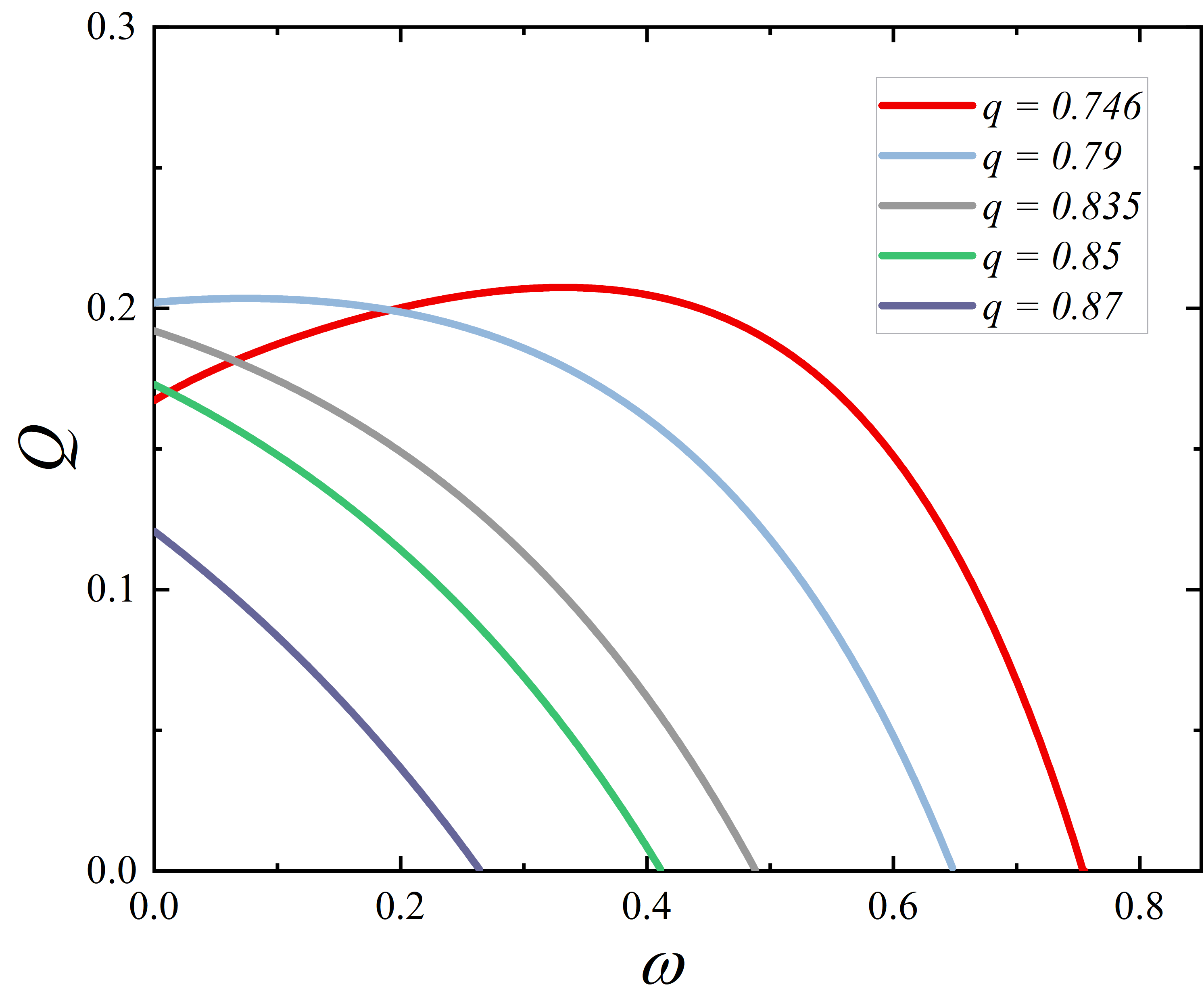}
			\label{fig:littlecharge}
		}

		\caption{The ADM mass $M$ and Noether charges $Q$ as a function of frequency with different magnetic charges $q$. In the top panel, starting at maximum frequency for BDSs solution with different magnetic charges and moving leftwards along the solutions curves, one encounters the first BDSs solution with the light ring at several solid geometric shapes (the dot, hexagon, and diamond).}
		\label{fig:matterandcharge}		
		\end{figure}

Considering the two global conserved charges of the solutions we obtained, namely the Noether charge $Q$ and the ADM mass $M$, we observe different behaviors in the curves of ADM mass $M$ or $Q$ versus $\omega$ for varying magnetic charges $q$. Firstly, when the magnetic charge $q<q_c=0.746$, these curves form a series of spirals. As shown in the left panel of Fig. \ref{fig:matterandcharge}, starting from $\omega=1$ for $M=0$ $(Q=0)$, as $\omega$ decreases for the first time, the mass  $M$ initially increases and then decreases, and we refer to this part of the solutions as the first branch. Subsequently, after reaching the minimum value of $\omega$, the mass continues to decrease, and the curve spirals back to form the second branch. We speculate that similar to general boson stars, Dirac stars, and Proca stars~\cite{Brito:2015pxa,Sanchis-Gual:2017bhw,Aoki:2022woy,Huang:2023glq,Liebling:2012fv}, if the calculations are extended further, additional branches may be discovered, ultimately leading to a central solution that approaches a critical singular solution. However, when the magnetic charge $q$ exceeds $0.746$, the frequency can approach zero, and the second branch will not appear.

\subsection{Light ring and light ball}
\begin{figure}[!htbp]
		\centering		
		\subfloat[]{ 
			\includegraphics[height=.30\textheight,width=.34\textheight, angle =0]{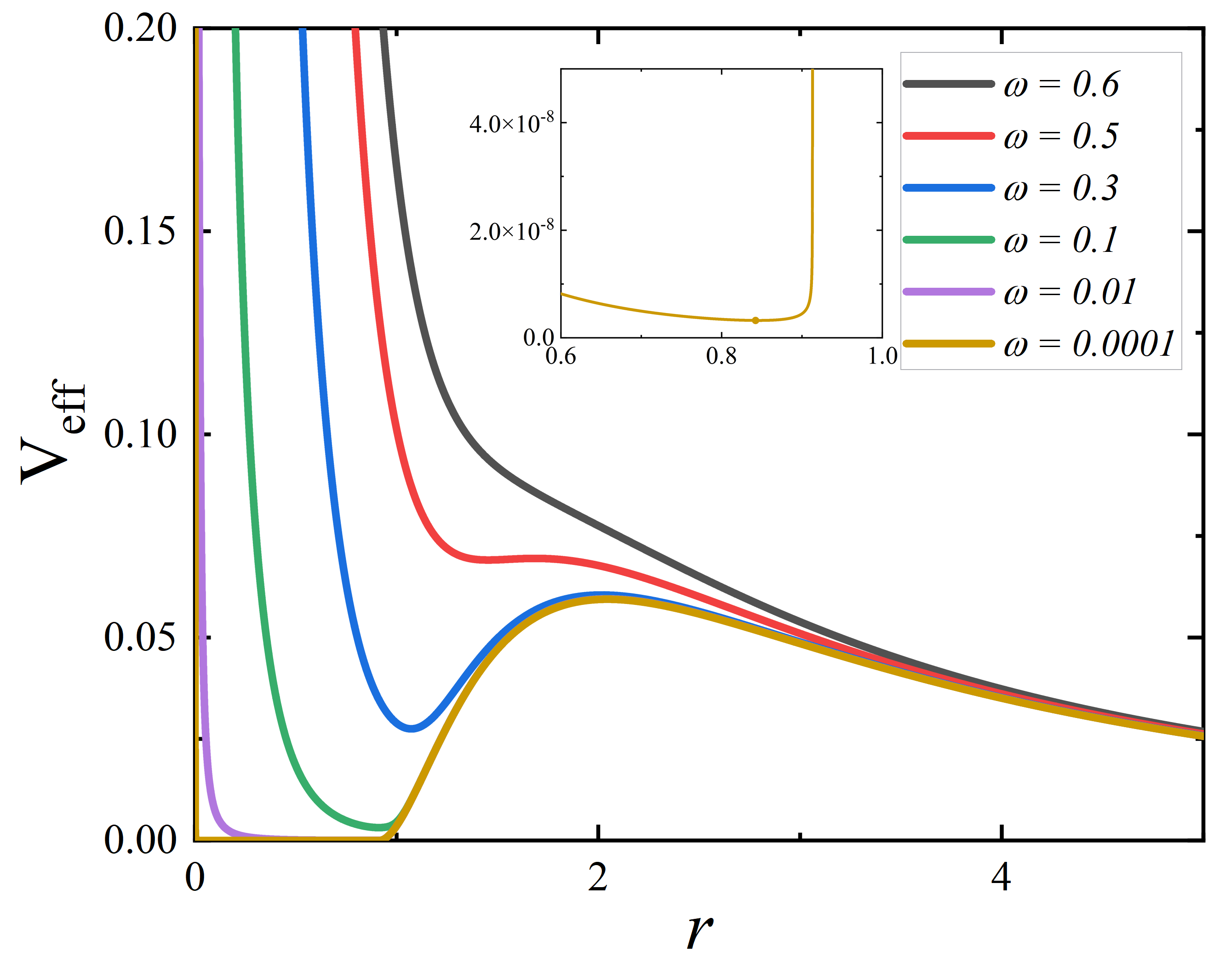}
			\label{fig:qeffective}
		}
       	\subfloat[]{
			\includegraphics[height=.30\textheight,width=.34\textheight, angle =0]{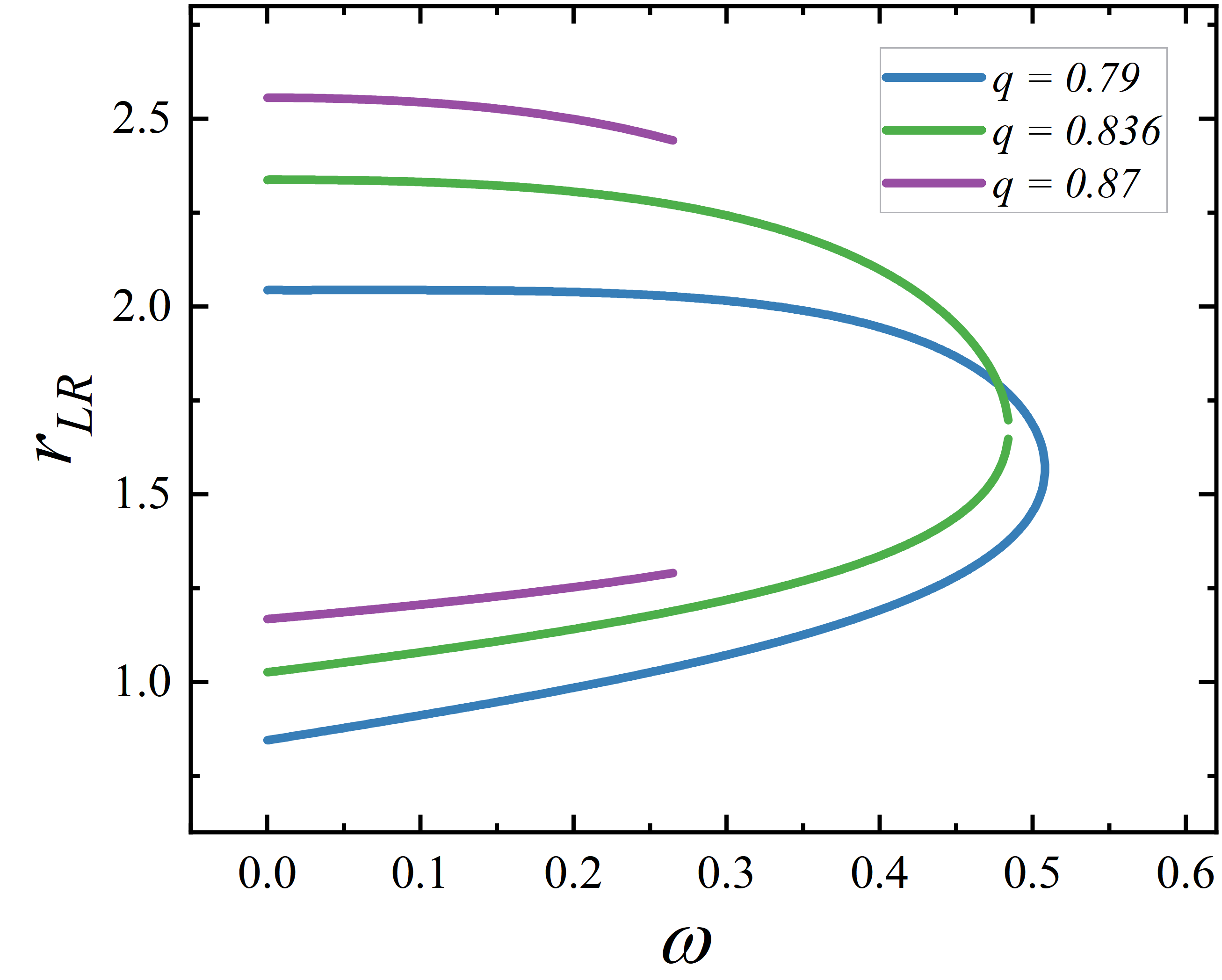}
			\label{fig:wrlr}
   		}
		\caption{Left panel: Effective potential as a function of radius $r$ for BDSs. Right panel: the relationship between the position of the light ring $r_{LR}$ and the frequency $\omega$ of Dirac fields for BDSs.}
		\label{fig:generaleffective}		
		\end{figure}

As shown in the top panel of Fig. \ref{fig:matterandcharge}, starting at the maximum frequency for BDSs solutions with different magnetic charges and moving leftwards along the solutions curves, we use various solid shapes to present the first BDSs solution with the light ring (LR). To better study the properties of the light rings, before discussing these light rings, we need to analyze the effective potential experienced by particles on null paths as they orbit around the BDSs. The expression for the effective potential can be derived through the geodesic equations for photons in a gravitational field, given by
\begin{equation}
    g_{\mu \nu} \Dot{x}^{\mu}\Dot{x}^{\nu}=0,
    \end{equation}
here, the dot represents the derivative with respect to the affine parameter $\lambda$ along the geodesic, and $x^{\mu}=(t,r,\theta,\varphi)$. Additionally, due to the static spherical symmetry, on the one hand, we can assume that the orbit lies in the equatorial plane ($\theta=\pi/2$). And on the other hand, the system possesses two Killing vectors $\partial_{t}$ and $\partial_{\varphi}$. Therefore, we can define the energy $E=-g_{tt}\dot{t}$ and the angular momentum $L=r^2 \dot{\varphi}$ of the null particle as two conserved quantities~\cite{Delgado:2021jxd}. Moreover, considering the ansatz of the metric (\ref{eq:metric}), consequently, we can obtain~\cite{Kar:2023dko,Rosa:2022tfv,Herdeiro:2021lwl}
\begin{equation}
    \Dot{r}^2 +\frac{1}{g_{tt}}\frac{1}{g_{rr}}\left(\frac{1}{b^2}+\frac{gtt}{r^2}\right)=\Dot{r}^2+\frac{1}{o^2}\left(\frac{no^2}{r^2}-\frac{1}{b^2}\right)=0,
\end{equation}
here, the impact parameter $b\equiv\frac{L}{E}$ is a constant for a certain null particle. We define the effective potential $V_{eff}$~\cite{Cardoso:2019rvt} is
\begin{equation}
    V_{eff}=\frac{no^2}{r^2}.
    \label{eq:effective}
\end{equation}
As an illustrative explanation, taking $q=0.85$ as an example, $V_{eff}$ of BDSs are plotted in Fig. \ref{fig:qeffective}. From the graph, it is shown that as the frequency decreases, the number of extreme points of the effective potential transitions from none to one and eventually to two. Where the extreme points $r_{LR}$ satisfy
\begin{equation}
    \frac{dV_{eff}}{dr}\bigg|_{r_{LR}} =0.
     \label{eq:effectivedaoshu}
\end{equation}
Therefore, for a null particle, if the square of the impact parameter $b^2_c=r^2_{LR}/n(r_{LR})o^2(r_{LR})$, when the particle arriving from elsewhere to radius $r_{LR}$, will have its radial velocity reduced to zero and remain at that location. The corresponding orbits of the null particle (i.e., $r=r_{LR}$) are then referred to as the light ring~\cite{Cardoso:2021sip}. Furthermore, at $r_{LR}$, 
 if the second derivative of the effective potential $V^{''}_{eff}(r_{LR})<0$, the light ring is unstable, manifested by the existence of a maximum point in the effective potential. Conversely, if the second derivative of the effective potential is greater than $0$, indicating the existence of a minimum point, then it corresponds to a stable light ring~\cite{Cunha:2022gde,Cunha:2017wao}.

Now we return to Fig. \ref{fig:matterandcharge}. It can be seen that when $q=0$ (i.e., the Dirac stars), starting at maximum frequency and moving along the solutions curves, the first BDSs solution with the light ring only appears on the second branch (green dot in the top left panel). However, if the magnetic charge $q=0.463$, the first solution with the light ring appears near the turning point of the first branch. We denote this with a brown dot. Subsequently, as $q$ continues to increase, the first BDSs solution with light ring shifts to the first branch, indicated by blue points. Starting from the blue dots and moving leftwards along the first branch solutions curves, there are two light rings. Until $q=0.835$ (blue hexagon), a light ring precisely appears near the maximum omega corresponding to the solution of BDSs. Afterward, if $q$ continues to increase, all solutions exhibit two light rings (represented by diamonds in the figure). Fig. \ref{fig:wrlr} further illustrates this process with the frequency $\omega$ vs. light ring position $r_{LR}$ relationship for BDSs with three different magnetic charges. When $q=0.79$, the positions of the innermost and outermost light rings approach each other as $\omega$ increases, eventually only a light ring remains. However, when $q=0.836>0.835$, it can be observed that the two curves representing the innermost light ring and outermost light ring no longer intersect. Subsequently, as $q$ continues to increase, the minimum distance between the two curves becomes increasingly larger.

After discussing the light ring of general BDSs, We focus on BDSs with lower frequencies, namely, FBDSs. To compare with the magnetic RN black hole, we normalize both the horizontal and vertical axes by dividing them by $M$ and use the radial coordinates $r$ before the conformal transformation. Where the metric for the RN black hole with magnetic charge $q$ and mass $M$ is given by
\begin{equation}
    ds^2=-\Delta(r)dt^2+\Delta(r)^{-1}(r)dr^2+r^2\left(d\theta^2+\sin^2 \theta d \varphi^2\right),
\end{equation}
where 
\begin{equation}
    \Delta(r)=1-\frac{2M}{r}+\frac{q^2}{r^2},
\end{equation}
the event horizon $r_H$ of magnetic RN black hole can be obtained from $\Delta=0$
\begin{equation}
    (r_H)_{\pm}=M\pm\sqrt{M^2-q^2}=M(1\pm\sqrt{1-\frac{q^2}{M^2}}).
\end{equation}
Especially, when the magnetic charge $q$ is zero, the RN black hole solution reduces to the Schwarzschild solution, and when $q^2=M^2$, it represents an extremal magnetic RN black hole solution.

	\begin{figure}[!htbp]
		\centering		
		\subfloat[]{
			\includegraphics[height=.30\textheight,width=.34\textheight, angle =0]{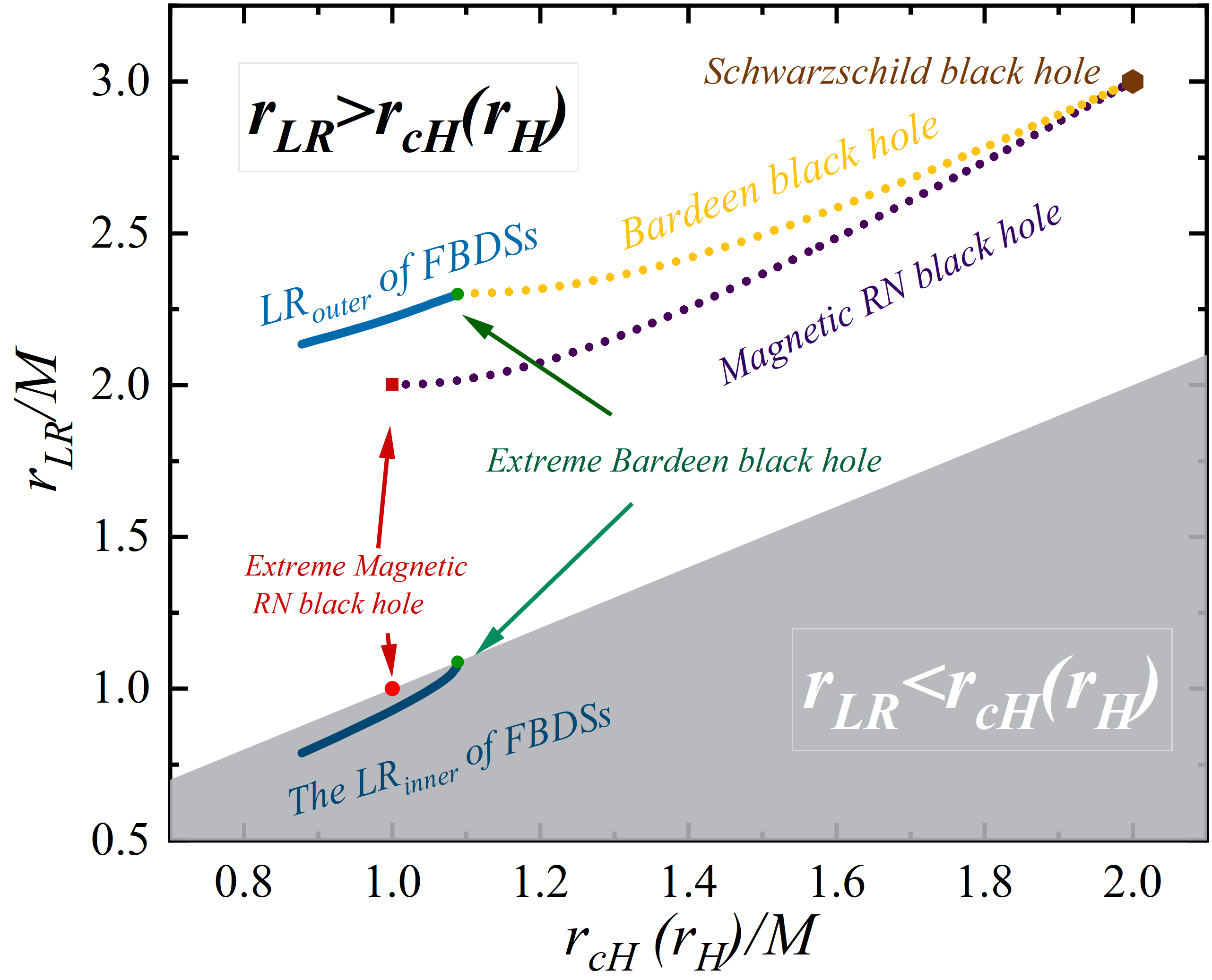}
   \label{fig:rhrl}
   		} 	
     		\subfloat[]{  
			\includegraphics[height=.30\textheight,width=.34\textheight, angle =0]{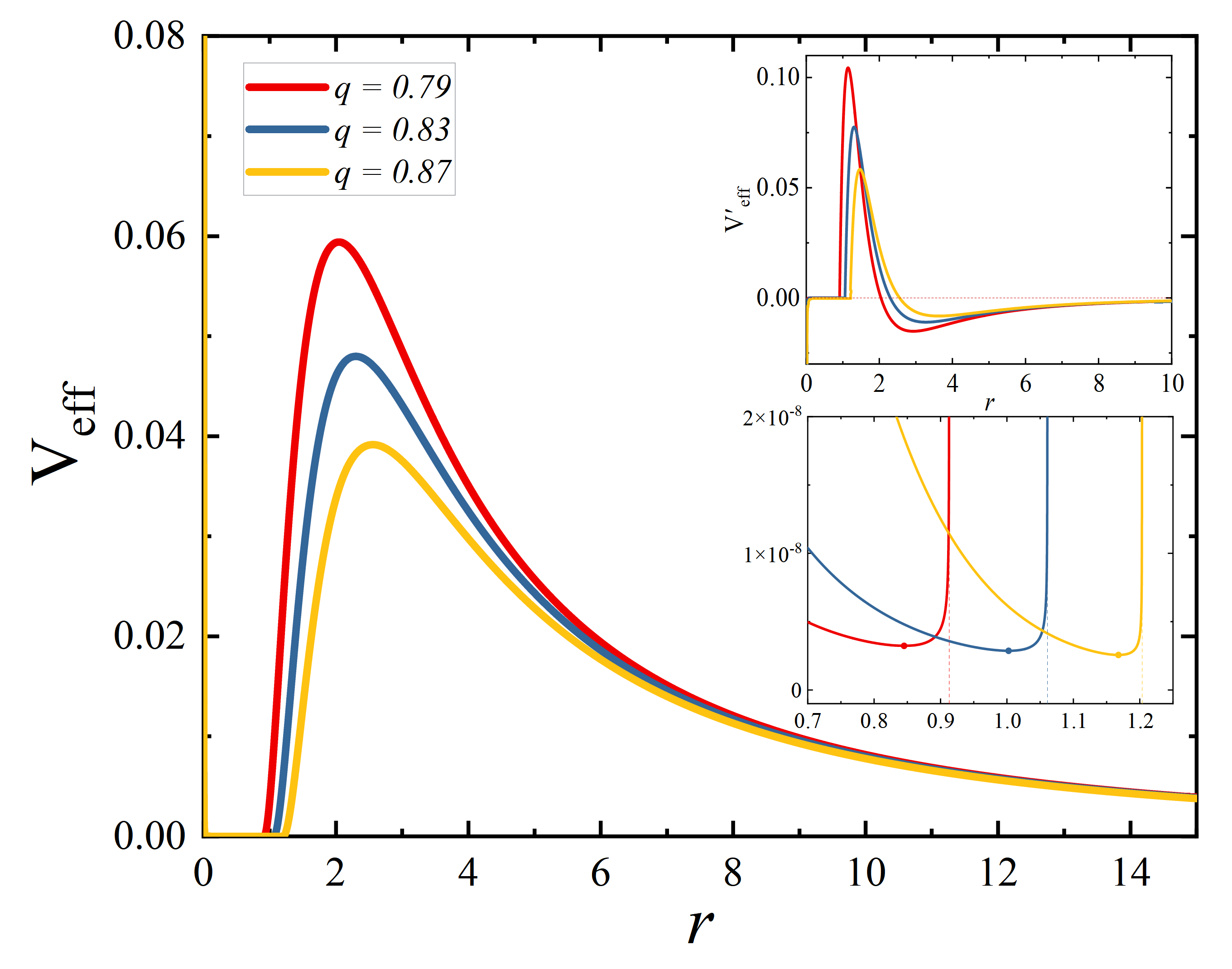}
      \label{fig:weffectiveanddaoshu}
		}
		\caption{Left panel: The relationship between the light ring positions $r_{LR}$ and the critical (event) horizon radius $r_{cH}$ ($r_{H}$) for different models. The gray region represents $r_{LR}<r_{cH} (r_H)$. Right panel: Effective potential and its derivative for three magnetic charges $q=0.79,0.83,0.87$ with fixed frequency $\omega=0.0001$.}
		\label{fig:feffective}		
		\end{figure}
It should be noted that since the spacetime inside the event horizon of a black hole is unobservable, we do not discuss the light ring inside the horizon. However, unlike the black hole, FBDSs have no event horizon, and there are observable effects within the critical horizon. Therefore, the light rings inside the critical horizon also need to be considered. For FBDSs and some other black holes, we provided the relation between the positions of the light ring $r_{LR}/M$ and the critical horizon radius $r_{cH}/M$ or event horizon radius $r_H/M$ in Fig. \ref{fig:rhrl}.

As shown in Fig. \ref{fig:rhrl}, firstly, it can be observed that the critical horizon radius of the FBDS is smaller than the extreme RN black hole that can achieve the smallest event horizon radius in several black hole models. Secondly, for the same $r_H(r_{cH})/M$, the outermost light ring of FBDSs is further away compared to other black holes, while the innermost light ring is closer to the origin. Specifically, when $r_{H}/M=1$, the position of the innermost light ring of FBDSs and extremal RN black holes (red dots) are very close. 

	\begin{figure}[!htbp]
		\centering		
		\subfloat{  
			\includegraphics[height=.20\textheight,width=.20\textheight, angle =0]{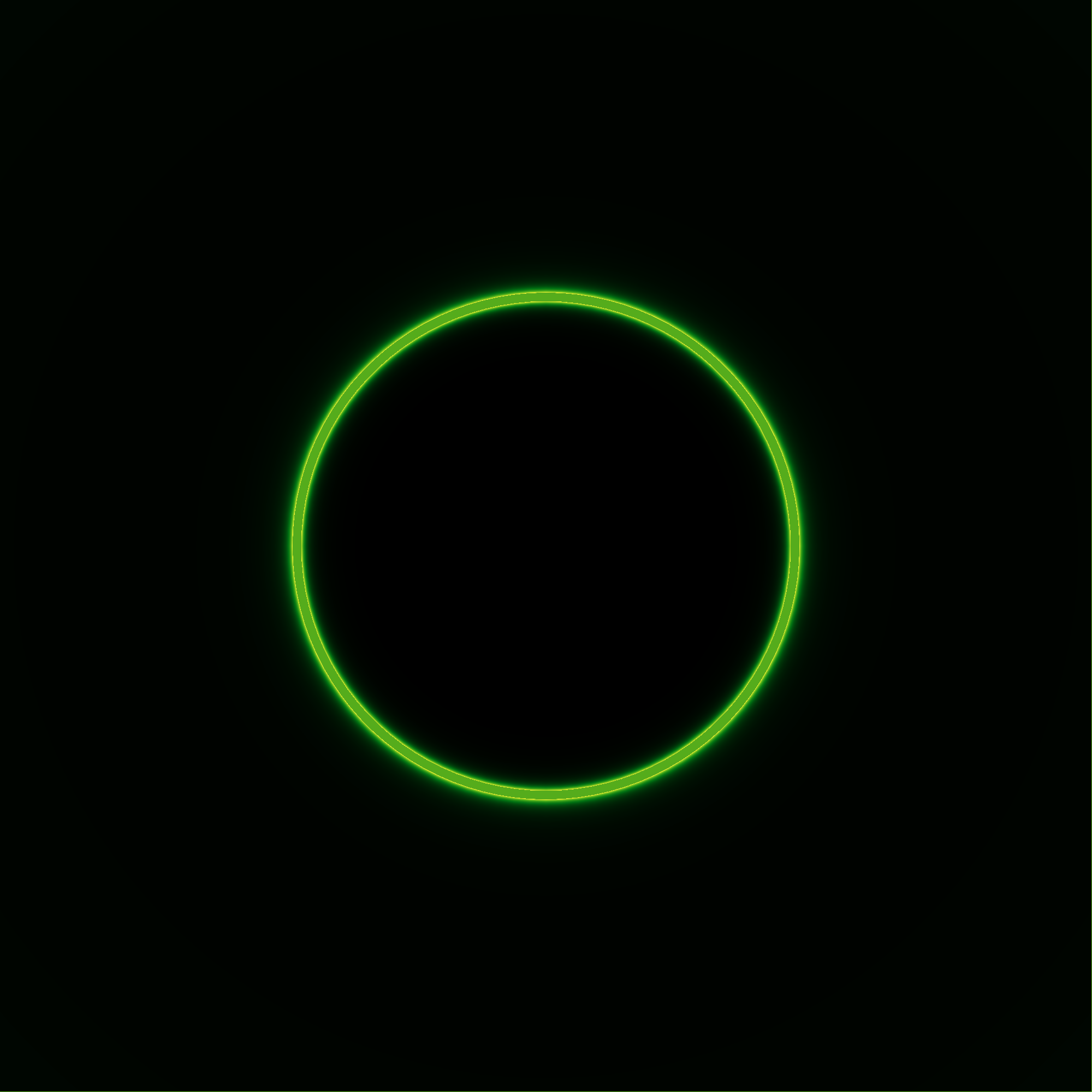}
		}
		\subfloat{
			\includegraphics[height=.20\textheight,width=.20\textheight, angle =0]{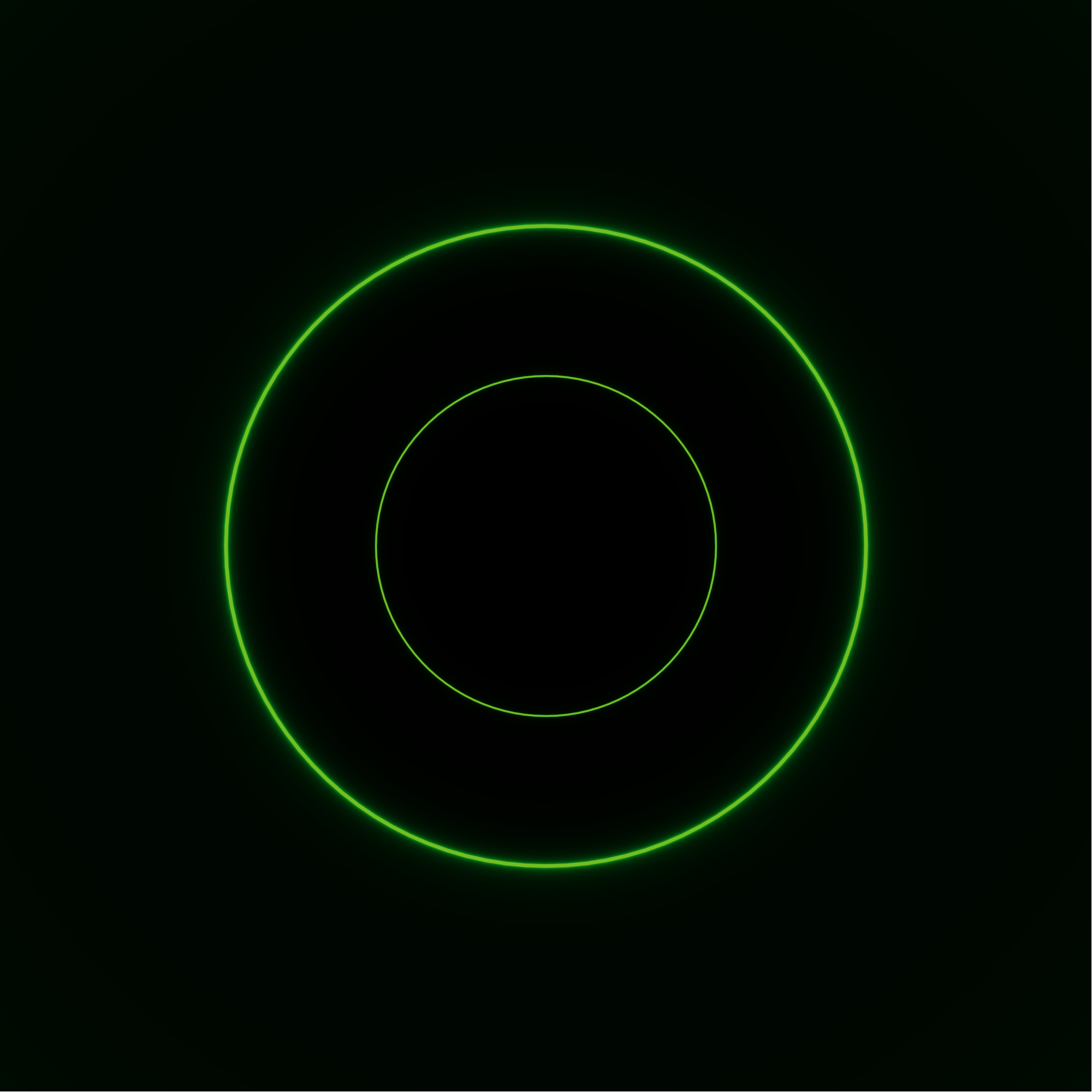}
   		} 	
     		\subfloat{
			\includegraphics[height=.20\textheight,width=.20\textheight, angle =0]{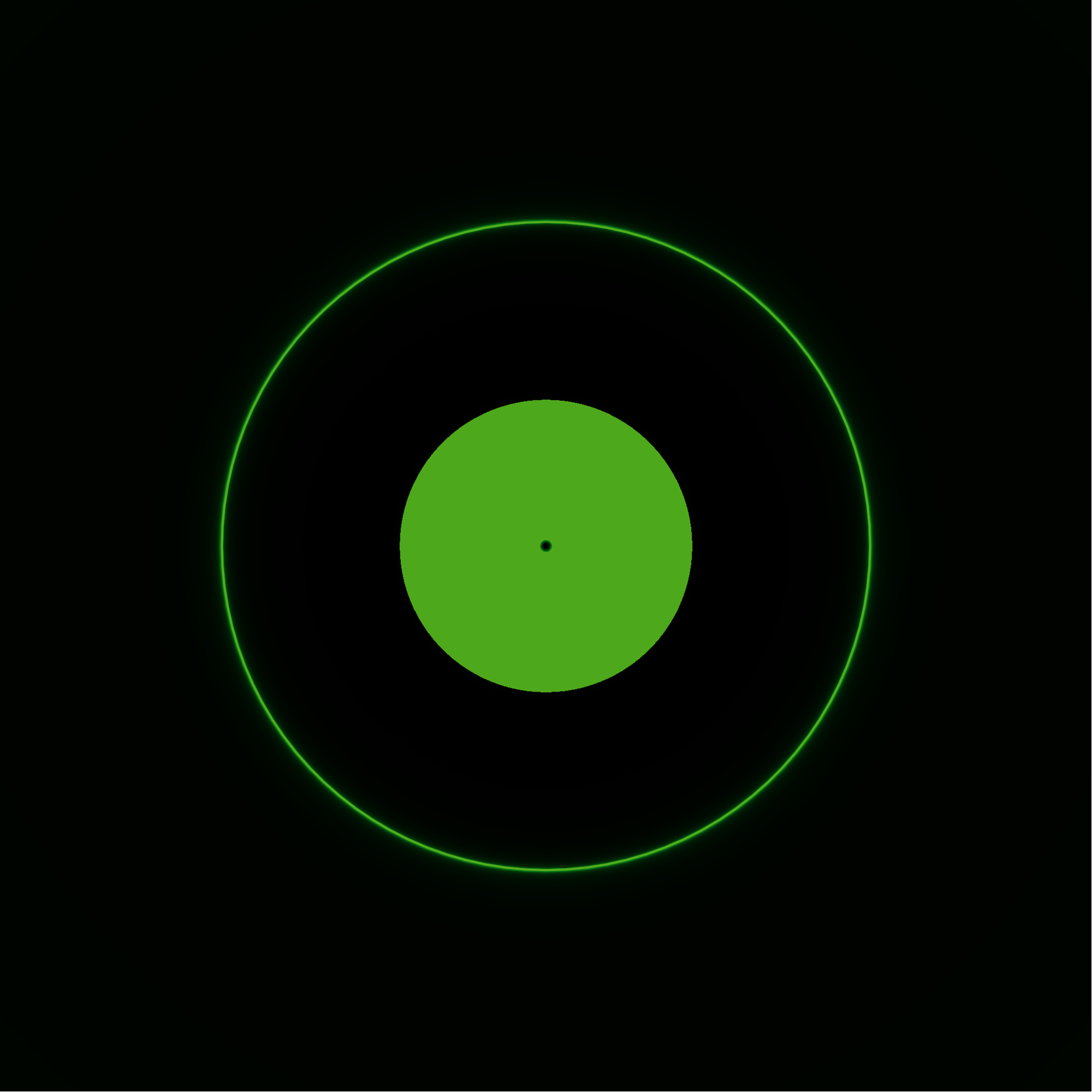}
   		} 	
		\caption{The illustrations picture of the light ring or light ball for three different Dirac field frequencies with $q=0.79$. From left to right panel, the frequencies are $0.508$, $0.3$, and $0.0001$, respectively.}
		\label{fig:lighringandball}		
		\end{figure}

However, it's worth noting that although there is strictly only one extreme point inside the critical horizon for effective potential, as shown in the bottom inset of Fig. \ref{fig:weffectiveanddaoshu}, inside the critical horizon, both the effective potential $V_{eff}$ and its rate of change $V^{\prime}_{eff}$  (the top inset of Fig. \ref{fig:weffectiveanddaoshu}) are extremely small, practically zero. Therefore, it can be considered that inside the $r_{cH}$, photons would take an extremely long time to reach the position of this light ring. Consequently, the photons form a ``light ball" instead of a ``light ring". Moreover, Fig. \ref{fig:lighringandball} provides an illustrative diagram using $q=0.79$ as an example for the light ring and light ball. When the frequency is relatively large (left panel), only one light ring appears. As the frequency decreases (middle panel), two light rings emerge, and finally, when the frequency approaches zero (right panel), the innermost ring transforms into a light ball. The shadow appearing at the origin in the right panel is due to the fact that, according to the formula (\ref{eq:effective}) and (\ref{eq:effectivedaoshu}), at the origin, the denominator strictly equals zero, while the numerator is merely approaching zero.  This causes the effective potential and its derivative for the null particle at the origin to be infinite. In other words, this implies that the velocity and the rate of change of velocity at the origin are infinite, indicating that there is no circular orbit at the origin.

\section{CONCLUSION}\label{sec: conclusion}
In this paper, we have investigated a model where Einstein's gravity is coupled with nonlinear electromagnetic fields and Dirac fields. We found that if the Dirac field is introduced, the magnetic charge of FBDSs is constrained within a certain range. Then for this range, we could not find black hole solutions. In other words, the introduction of the Dirac field prevents the Bardeen spacetime from possessing black hole solutions. Moreover, when the magnetic charge is greater than $0.746$ and the frequency of the Dirac field approaches zero, we found the FBDSs solutions. Outside the critical horizon, these FBDSs solutions manifest spacetime properties remarkably similar to those of extreme black holes.

Using the effective potential, we found that FBDSs have two light rings, with one residing on each side of the critical horizon. The outermost light ring is unstable and indeed is a ring. However, for the stable innermost light ring, since the metric function $no^2$ is very close to zero inside the $r_{cH}$, the effective potential and its derivative are very little. This results in the photons requiring a considerable long time to traverse a short distance. Therefore, the interior of the critical horizon is practically a light ball, rather than a light ring.

There are many interesting extensions of our work. On the one hand, if FBDSs exhibit Hawking radiation, due to the absence of an event horizon in their solutions preventing the occurrence of the information loss paradox associated with Hawking radiation, studying Hawking radiation in such models is highly worthwhile. On the other hand, one would wonder whether our model can be stable against small perturbations. Therefore, the dynamical stability of FBDSs is also a crucial subject for further investigation.

\section*{ACKNOWLEDGEMENTS}
	This work is supported by National Key Research and Development Program of China (Grant No. 2020YFC2201503) and the National Natural Science Foundation of China (Grants No.~12275110 and No.~12047501).


\begin{thebibliography}{99}
	
    \bibitem{KAGRA:2021duu}
    R.~Abbott \textit{et al.} [KAGRA, VIRGO and LIGO Scientific],
    ``Population of Merging Compact Binaries Inferred Using Gravitational Waves through GWTC-3,''
    Phys. Rev. X \textbf{13} (2023) no.1, 011048
    [arXiv:2111.03634 [astro-ph.HE]].
%
    \bibitem{Ghez:2003qj}
    A.~M.~Ghez, S.~Salim, S.~D.~Hornstein, A.~Tanner, M.~Morris, E.~E.~Becklin and G.~Duchene,
    ``Stellar orbits around the galactic center black hole,''
    Astrophys. J. \textbf{620} (2005), 744-757
    [arXiv:astro-ph/0306130 [astro-ph]].
    \bibitem{Ghez:2008ms}
    A.~M.~Ghez, S.~Salim, N.~N.~Weinberg, J.~R.~Lu, T.~Do, J.~K.~Dunn, K.~Matthews, M.~Morris, S.~Yelda and E.~E.~Becklin, \textit{et al.}
    ``Measuring Distance and Properties of the Milky Way's Central Supermassive Black Hole with Stellar Orbits,''
    Astrophys. J. \textbf{689} (2008), 1044-1062
    [arXiv:0808.2870 [astro-ph]].
%
    \bibitem{Rhoades:1974fn}
    C.~E.~Rhoades, Jr. and R.~Ruffini,
    ``Maximum mass of a neutron star,''
    Phys. Rev. Lett. \textbf{32} (1974), 324-327
    \bibitem{Kalogera:1996ci}
    V.~Kalogera and G.~Baym,
    ``The maximum mass of a neutron star,''
    Astrophys. J. Lett. \textbf{470} (1996), L61-L64
    [arXiv:astro-ph/9608059 [astro-ph]].
    
\bibitem{NANOGrav:2019jur}
H.~T.~Cromartie \textit{et al.} [NANOGrav],
``Relativistic Shapiro delay measurements of an extremely massive millisecond pulsar,''
Nature Astron. \textbf{4} (2019) no.1, 72-76
[arXiv:1904.06759 [astro-ph.HE]].
    %
\bibitem{Kunz:2022wnj}
J.~Kunz,
``Neutron Stars,''
Lect. Notes Phys. \textbf{1017} (2023), 293-313
[arXiv:2204.12520 [gr-qc]].
    \bibitem{Herdeiro:2015gia}
    C.~Herdeiro and E.~Radu,
    ``Construction and physical properties of Kerr black holes with scalar hair,''
    Class. Quant. Grav. \textbf{32} (2015) no.14, 144001
    [arXiv:1501.04319 [gr-qc]].
    \bibitem{LIGOScientific:2016aoc}
    B.~P.~Abbott \textit{et al.} [LIGO Scientific and Virgo],
    ``Observation of Gravitational Waves from a Binary Black Hole Merger,''
    Phys. Rev. Lett. \textbf{116} (2016) no.6, 061102
    [arXiv:1602.03837 [gr-qc]].
    \bibitem{Falcke:1999pj}
    H.~Falcke, F.~Melia and E.~Agol,
    ``Viewing the shadow of the black hole at the galactic center,''
    Astrophys. J. Lett. \textbf{528} (2000), L13
    [arXiv:astro-ph/9912263 [astro-ph]].


    
    \bibitem{EventHorizonTelescope:2019dse}
    K.~Akiyama \textit{et al.} [Event Horizon Telescope],
    ``First M87 Event Horizon Telescope Results. I. The Shadow of the Supermassive Black Hole,''
    Astrophys. J. Lett. \textbf{875} (2019), L1
    [arXiv:1906.11238 [astro-ph.GA]].
    \bibitem{EventHorizonTelescope:2019uob}
    K.~Akiyama \textit{et al.} [Event Horizon Telescope],
    ``First M87 Event Horizon Telescope Results. II. Array and Instrumentation,''
    Astrophys. J. Lett. \textbf{875} (2019) no.1, L2
    [arXiv:1906.11239 [astro-ph.IM]].
    \bibitem{EventHorizonTelescope:2019jan}
    K.~Akiyama \textit{et al.} [Event Horizon Telescope],
    ``First M87 Event Horizon Telescope Results. III. Data Processing and Calibration,''
    Astrophys. J. Lett. \textbf{875} (2019) no.1, L3
    [arXiv:1906.11240 [astro-ph.GA]].

    \bibitem{EventHorizonTelescope:2019ths}
    K.~Akiyama \textit{et al.} [Event Horizon Telescope],
    ``First M87 Event Horizon Telescope Results. IV. Imaging the Central Supermassive Black Hole,''
    Astrophys. J. Lett. \textbf{875} (2019) no.1, L4
    [arXiv:1906.11241 [astro-ph.GA]].
    \bibitem{EventHorizonTelescope:2019pgp}
    K.~Akiyama \textit{et al.} [Event Horizon Telescope],
    ``First M87 Event Horizon Telescope Results. V. Physical Origin of the Asymmetric Ring,''
    Astrophys. J. Lett. \textbf{875} (2019) no.1, L5
    [arXiv:1906.11242 [astro-ph.GA]].
    \bibitem{EventHorizonTelescope:2019ggy}
    K.~Akiyama \textit{et al.} [Event Horizon Telescope],
    ``First M87 Event Horizon Telescope Results. VI. The Shadow and Mass of the Central Black Hole,''
    Astrophys. J. Lett. \textbf{875} (2019) no.1, L6
    [arXiv:1906.11243 [astro-ph.GA]].
    \bibitem{Einstein:1939ms}
    A.~Einstein,
    ``On a stationary system with spherical symmetry consisting of many gravitating masses,''
    Annals Math. \textbf{40} (1939), 922-936

    \bibitem{Lan:2023cvz}
    C.~Lan, H.~Yang, Y.~Guo and Y.~G.~Miao,
    ``Regular Black Holes: A Short Topic Review,''
    Int. J. Theor. Phys. \textbf{62} (2023) no.9, 202
    [arXiv:2303.11696 [gr-qc]].
    \bibitem{Sakharov:1966aja}
    A.~D.~Sakharov,
    ``Nachal'naia stadija rasshirenija Vselennoj i vozniknovenije neodnorodnosti raspredelenija veshchestva,''
    Sov. Phys. JETP \textbf{22} (1966), 241

    \bibitem{Gliner:1966}
    E.~B.~Gliner,
    ``Algebraic Properties of the Energy-momentum Tensor and Vacuum-like States of Matter,''
    Sov. Phys. JETP \textbf{22} (1966), 378

    

    \bibitem{Bardeen:1968}
    J.~Bardeen,
    in Proceedings of GR5,
    Tiflis, U.S.S.R. (1968).
    [arXiv:2303.11696 [gr-qc]].
    \bibitem{Ayon-Beato:1998hmi}
    E.~Ayon-Beato and A.~Garcia,
    ``Regular black hole in general relativity coupled to nonlinear electrodynamics,''
    Phys. Rev. Lett. \textbf{80} (1998), 5056-5059
    [arXiv:gr-qc/9911046 [gr-qc]].

\bibitem{Wheeler:1955zz}
J.~A.~Wheeler,
``Geons,''
Phys. Rev. \textbf{97} (1955), 511-536
\bibitem{Schunck:2003kk}
F.~E.~Schunck and E.~W.~Mielke,
``General relativistic boson stars,''
Class. Quant. Grav. \textbf{20} (2003), R301-R356
[arXiv:0801.0307 [astro-ph]].
\bibitem{Liebling:2012fv}
S.~L.~Liebling and C.~Palenzuela,
``Dynamical boson stars,''
Living Rev. Rel. \textbf{26} (2023) no.1, 1
[arXiv:1202.5809 [gr-qc]].
\bibitem{Herdeiro:2021mol}
C.~A.~R.~Herdeiro, J.~Kunz, I.~Perapechka, E.~Radu and Y.~Shnir,
``Chains of Boson Stars,''
Phys. Rev. D \textbf{103} (2021) no.6, 065009
[arXiv:2101.06442 [gr-qc]].
\bibitem{CalderonBustillo:2020fyi}
J.~Calder\'on Bustillo, N.~Sanchis-Gual, A.~Torres-Forn\'e, J.~A.~Font, A.~Vajpeyi, R.~Smith, C.~Herdeiro, E.~Radu and S.~H.~W.~Leong,
``GW190521 as a Merger of Proca Stars: A Potential New Vector Boson of $8.7\times 10^{-13}$  eV,''
Phys. Rev. Lett. \textbf{126} (2021) no.8, 081101
[arXiv:2009.05376 [gr-qc]].
\bibitem{CalderonBustillo:2022cja}
J.~Calderon Bustillo, N.~Sanchis-Gual, S.~H.~W.~Leong, K.~Chandra, A.~Torres-Forne, J.~A.~Font, C.~Herdeiro, E.~Radu, I.~C.~F.~Wong and T.~G.~F.~Li,
``Searching for vector boson-star mergers within LIGO-Virgo intermediate-mass black-hole merger candidates,''
[arXiv:2206.02551 [gr-qc]].
\bibitem{Lee:1995af}
J.~w.~Lee and I.~g.~Koh,
``Galactic halos as boson stars,''
Phys. Rev. D \textbf{53} (1996), 2236-2239
[arXiv:hep-ph/9507385 [hep-ph]].
\bibitem{Suarez:2013iw}
A.~Su\'arez, V.~H.~Robles and T.~Matos,
``A Review on the Scalar Field/Bose-Einstein Condensate Dark Matter Model,''
Astrophys. Space Sci. Proc. \textbf{38} (2014), 107-142
[arXiv:1302.0903 [astro-ph.CO]].
\bibitem{Eby:2015hsq}
J.~Eby, C.~Kouvaris, N.~G.~Nielsen and L.~C.~R.~Wijewardhana,
``Boson Stars from Self-Interacting Dark Matter,''
JHEP \textbf{02} (2016), 028
[arXiv:1511.04474 [hep-ph]].
\bibitem{Chen:2020cef}
J.~Chen, X.~Du, E.~W.~Lentz, D.~J.~E.~Marsh and J.~C.~Niemeyer,
``New insights into the formation and growth of boson stars in dark matter halos,''
Phys. Rev. D \textbf{104} (2021) no.8, 083022
[arXiv:2011.01333 [astro-ph.CO]].
    \bibitem{Wang:2023tdz}
    X.~E.~Wang,
    ``From Bardeen-boson stars to black holes without event horizon,''
    [arXiv:2305.19057 [gr-qc]].
\bibitem{zeldovichbookorpaper} Y. B. Zel'dovich and I. D. Novikov,
{\it Relativistic Astrophysics 1: Stars and Relativity}, (University
of Chicago Press, Chicago 1971), p. 369 (translation from the 1967
Russian edition).



\bibitem{Ruffini:1971bza}
R.~Ruffini and J.~A.~Wheeler,
``Introducing the black hole,''
Phys. Today \textbf{24} (1971) no.1, 30
\bibitem{Vachaspati:2007fc}
T.~Vachaspati,
``Black Stars and Gamma Ray Bursts,''
[arXiv:0706.1203 [astro-ph]].
\bibitem{Schmelzer:2010zq}
I.~Schmelzer,
``Black Holes or Frozen Stars? A Viable Theory of Gravity without Black Holes,''
[arXiv:1003.1446 [physics.gen-ph]].
\bibitem{Lemos:2010te}
J.~P.~S.~Lemos and V.~T.~Zanchin,
``Quasiblack holes with pressure: relativistic charged spheres as the frozen stars,''
Phys. Rev. D \textbf{81} (2010), 124016
[arXiv:1004.3574 [gr-qc]].

\bibitem{Zhang:2010vh}
S.~N.~Zhang,
``On the Solution to the 'Frozen Star' Paradox, Nature of Astrophysical Black Holes, non-Existence of Gravitational Singularity in the Physical Universe and Applicability of the Birkhoff's Theorem,''
Int. J. Mod. Phys. D \textbf{20} (2011), 1891-1899
[arXiv:1003.1359 [gr-qc]].
\bibitem{Vachaspati:2016hya}
T.~Vachaspati,
``Gravitational Waves, Gamma Ray Bursts, and Black Stars,''
Int. J. Mod. Phys. D \textbf{25} (2016) no.12, 1644025
[arXiv:1611.03853 [gr-qc]].
\bibitem{Brustein:2023hic}
R.~Brustein, A.~J.~M.~Medved and T.~Simhon,
``Thermodynamics of frozen stars,''
[arXiv:2310.11572 [gr-qc]].
\bibitem{Zeng:2023ueq}
D.~f.~Zeng,
``Microscopic State of BHs and an Exact One Body Method for Binary Dynamics in General Relativity,''
[arXiv:2311.11764 [gr-qc]].
\bibitem{Kastor:2016cqs}
D.~Kastor and J.~Traschen,
``Building Cosmological Frozen Stars,''
Class. Quant. Grav. \textbf{34} (2017) no.3, 035012
[arXiv:1607.08219 [hep-th]].
\bibitem{Leith:2020jqw}
P.~E.~D.~Leith, C.~A.~Hooley, K.~Horne and D.~G.~Dritschel,
``Fermion self-trapping in the optical geometry of Einstein-Dirac solitons,''
Phys. Rev. D \textbf{101} (2020) no.10, 106012
[arXiv:2002.02747 [gr-qc]].
\bibitem{Leith:2021urf}
P.~E.~D.~Leith, C.~A.~Hooley, K.~Horne and D.~G.~Dritschel,
``Nonlinear effects in the excited states of many-fermion Einstein-Dirac solitons,''
Phys. Rev. D \textbf{104} (2021) no.4, 046024
[arXiv:2105.12672 [gr-qc]].
\bibitem{Finster:1998ws}
F.~Finster, J.~Smoller and S.~T.~Yau,
``Particle - like solutions of the Einstein-Dirac equations,''
Phys. Rev. D \textbf{59} (1999), 104020
[arXiv:gr-qc/9801079 [gr-qc]].
\bibitem{Dzhunushaliev:2018jhj}
V.~Dzhunushaliev and V.~Folomeev,
``Dirac stars supported by nonlinear spinor fields,''
Phys. Rev. D \textbf{99} (2019) no.8, 084030
[arXiv:1811.07500 [gr-qc]].
\bibitem{Dzhunushaliev:2019kiy}
V.~Dzhunushaliev and V.~Folomeev,
``Dirac star in the presence of Maxwell and Proca fields,''
Phys. Rev. D \textbf{99} (2019) no.10, 104066
[arXiv:1901.09905 [gr-qc]].
\bibitem{Dzhunushaliev:2019uft}
V.~Dzhunushaliev and V.~Folomeev,
``Dirac Star with SU(2) Yang-Mills and Proca Fields,''
Phys. Rev. D \textbf{101} (2020) no.2, 024023
[arXiv:1911.11614 [gr-qc]].
\bibitem{Leith:2022icf}
P.~E.~D.~Leith, A.~D.~Leggat, C.~A.~Hooley, K.~Horne and D.~G.~Dritschel,
``Gravitationally localized states of two neutral fermions interacting with a Higgs field,''
Phys. Rev. D \textbf{107} (2023) no.10, 106020
[arXiv:2202.03228 [gr-qc]].
\bibitem{Dzhunushaliev:2022wnd}
V.~Dzhunushaliev, V.~Folomeev and N.~Burtebayev,
``Rapidly rotating Dirac stars,''
Phys. Rev. D \textbf{106} (2022) no.2, 024021
[arXiv:2205.08707 [gr-qc]].

\bibitem{Liang:2023ywv}
C.~Liang, S.~X.~Sun, J.~R.~Rena and Y.~Q.~Wang,
``Multi-state Dirac stars,''
[arXiv:2306.11437 [hep-th]].
\bibitem{Sun:2023bdh}
S.~X.~Sun, S.~Y.~Cui, L.~X.~Huang and Y.~Q.~Wang,
``$\kappa$-Dirac stars,''
[arXiv:2310.10267 [gr-qc]].
\bibitem{Herdeiro:2017fhv}
C.~A.~R.~Herdeiro, A.~M.~Pombo and E.~Radu,
``Asymptotically flat scalar, Dirac and Proca stars: discrete vs. continuous families of solutions,''
Phys. Lett. B \textbf{773} (2017), 654-662
[arXiv:1708.05674 [gr-qc]].
\bibitem{Herdeiro:2019mbz}
C.~Herdeiro, I.~Perapechka, E.~Radu and Y.~Shnir,
``Asymptotically flat spinning scalar, Dirac and Proca stars,''
Phys. Lett. B \textbf{797} (2019), 134845
[arXiv:1906.05386 [gr-qc]].
\bibitem{Herdeiro:2020jzx}
C.~A.~R.~Herdeiro and E.~Radu,
``Asymptotically flat, spherical, self-interacting scalar, Dirac and Proca stars,''
Symmetry \textbf{12} (2020) no.12, 2032
[arXiv:2012.03595 [gr-qc]].
\bibitem{Ayon-Beato:2000mjt}
E.~Ayon-Beato and A.~Garcia,
``The Bardeen model as a nonlinear magnetic monopole,''
Phys. Lett. B \textbf{493} (2000), 149-152
[arXiv:gr-qc/0009077 [gr-qc]].

\bibitem{Dolan:2015eua}
S.~R.~Dolan and D.~Dempsey,
``Bound states of the Dirac equation on Kerr spacetime,''
Class. Quant. Grav. \textbf{32} (2015) no.18, 184001
[arXiv:1504.03190 [gr-qc]].
\bibitem{Brito:2015pxa}
R.~Brito, V.~Cardoso, C.~A.~R.~Herdeiro and E.~Radu,
``Proca stars: Gravitating Bose\textendash{}Einstein condensates of massive spin 1 particles,''
Phys. Lett. B \textbf{752} (2016), 291-295
[arXiv:1508.05395 [gr-qc]].
\bibitem{Sanchis-Gual:2017bhw}
N.~Sanchis-Gual, C.~Herdeiro, E.~Radu, J.~C.~Degollado and J.~A.~Font,
``Numerical evolutions of spherical Proca stars,''
Phys. Rev. D \textbf{95} (2017) no.10, 104028
[arXiv:1702.04532 [gr-qc]].
\bibitem{Aoki:2022woy}
K.~Aoki and M.~Minamitsuji,
``Resolving the pathologies of self-interacting Proca fields: A case study of Proca stars,''
Phys. Rev. D \textbf{106} (2022) no.8, 084022
[arXiv:2206.14320 [gr-qc]].
\bibitem{Huang:2023glq}
L.~X.~Huang, S.~X.~Sun, R.~Zhang, C.~Liang and Y.~Q.~Wang,
``Excited Dirac stars with higher azimuthal harmonic index,''
[arXiv:2309.16497 [gr-qc]].
\bibitem{Delgado:2021jxd}
J.~F.~M.~Delgado, C.~A.~R.~Herdeiro and E.~Radu,
``Equatorial timelike circular orbits around generic ultracompact objects,''
Phys. Rev. D \textbf{105} (2022) no.6, 064026
doi:10.1103/PhysRevD.105.064026
[arXiv:2107.03404 [gr-qc]].
\bibitem{Kar:2023dko}
A.~Kar and S.~Kar,
``Novel regular black holes: geometry, source and shadow,''
[arXiv:2308.12155 [gr-qc]].
\bibitem{Rosa:2022tfv}
J.~L.~Rosa and D.~Rubiera-Garcia,
``Shadows of boson and Proca stars with thin accretion disks,''
Phys. Rev. D \textbf{106} (2022) no.8, 084004
[arXiv:2204.12949 [gr-qc]].
\bibitem{Herdeiro:2021lwl}
C.~A.~R.~Herdeiro, A.~M.~Pombo, E.~Radu, P.~Cunha, V.P. and N.~Sanchis-Gual,
``The imitation game: Proca stars that can mimic the Schwarzschild shadow,''
JCAP \textbf{04} (2021), 051
[arXiv:2102.01703 [gr-qc]].


\bibitem{Cardoso:2019rvt}
V.~Cardoso and P.~Pani,
Living Rev. Rel. \textbf{22} (2019) no.1, 4
doi:10.1007/s41114-019-0020-4
[arXiv:1904.05363 [gr-qc]].
\bibitem{Cardoso:2021sip}
V.~Cardoso, F.~Duque and A.~Foschi,
``Light ring and the appearance of matter accreted by black holes,''
Phys. Rev. D \textbf{103} (2021) no.10, 104044
[arXiv:2102.07784 [gr-qc]].

\bibitem{Cunha:2022gde}
P.~V.~P.~Cunha, C.~Herdeiro, E.~Radu and N.~Sanchis-Gual,
``Exotic Compact Objects and the Fate of the Light-Ring Instability,''
Phys. Rev. Lett. \textbf{130} (2023) no.6, 061401
[arXiv:2207.13713 [gr-qc]].
\bibitem{Cunha:2017wao}
P.~V.~P.~Cunha, J.~A.~Font, C.~Herdeiro, E.~Radu, N.~Sanchis-Gual and M.~Zilh\~ao,
``Lensing and dynamics of ultracompact bosonic stars,''
Phys. Rev. D \textbf{96} (2017) no.10, 104040
[arXiv:1709.06118 [gr-qc]].

	


\end{thebibliography}
\end{document}